\documentclass{jfm}
\usepackage{amsmath}
\usepackage{epsfig,psfrag}
\usepackage{amsfonts}
\usepackage{mathrsfs}
\usepackage{amssymb}
\usepackage{bm}
\usepackage{color}
\usepackage{natbib}
\usepackage{mdwlist}
\usepackage{xr}

\newcommand\beq{\begin{equation}}
\newcommand\eeq{\end{equation}}
\newcommand\beqa{\begin{eqnarray}}
\newcommand\eeqa{\end{eqnarray}}
\newcommand{\cl}{\mathcal{L}}

\def\half{\frac{1}{2}}

\def\d{{\rm d}}

\shorttitle{Water wave precession resonance}
\shortauthor{Lucas, Bustamante \& Perlin}

\title{Precession resonance in water waves}
\author{Dan Lucas\aff{1}, Miguel D. Bustamante\aff{2}\corresp{\email{miguel.bustamante@ucd.ie}} \and Marc Perlin\aff{3}}
\affiliation
{
\aff{1}
{DAMTP, University of Cambridge, Cambridge, CB3 0WA, UK}
\aff{2}
{Institute for Discovery, School of Mathematics and Statistics, University College Dublin, Belfield, Dublin 4, Ireland}
\aff{3}
{NAME, University of Michigan, Ann Arbor, MI48108, USA}
}

\begin{document}

\maketitle

\begin{abstract}

We describe the theory and present numerical evidence for a new type of nonlinear resonant interaction between gravity waves on the surface of deep water. The resonance constitutes a generalisation of the usual `exact' resonance as we show that exchanges of energy between the waves can be enhanced when the interaction is three-wave rather than four and the linear frequency mismatch, or detuning, is non-zero i.e. $\omega_1\pm\omega_2\pm\omega_3 \neq0.$ This is possible because the resonance condition is now a match between the so-called `precession frequency' of a given \emph{{triad} interaction} and an existent nonlinear frequency in the system. In the limit of weak nonlinearity this precession frequency is simply due to the linear `drift' of the triad phase; therefore, it tends toward the detuning. This means precession resonance of this type can occur at finite amplitudes, with nonlinear corrections contributing to the resonance. {We report energy transfer efficiencies of up to 40\%, 
depending on the model options.} To the authors' knowledge this represents the first new type of nonlinear resonance in surface gravity waves since the seminal work of \cite{1967JFM....27..417B}. 
\end{abstract}


\section{Introduction}

Historically, nonlinear resonant interactions in surface water waves have focused mainly on the so-called exact resonances, defined by the equations
$$k_1 \pm ... \pm k_N = 0, \qquad \quad \omega_1 \pm ... \pm \omega_N = 0,$$
 where $N$ denotes the number of interacting waves, $k_j$ denote the wave-vectors, and $\omega_j$ denote the frequencies where $\omega_j = \omega(k_j)$ is provided by a dispersion relation. Early theoretical developments for these interactions gave rise to \emph{resonance interaction theory}, originated by Phillips (1960) and developed by several authors (see the reviews by \cite{HAMMACK:1993ht} {and \cite{nazarenko2016wave}}, and references therein). The key working hypothesis of this theory is the smallness of the wave steepness, namely the product between wave amplitude and magnitude of the wave-vector. A formal perturbation theory based on multiple-scale methods is thus developed which allows one to obtain the exact resonances as necessary conditions to avoid secular behaviour. Solutions to these exact resonance equations can be found in many cases. Case $N=3$, called triad resonances, applies to capillary-gravity waves (and also to a variety of wave systems such as Charney-Hasegawa-Mima, quasi-geostrophic equations, internal waves, inertial waves, etc.); case $N=4$, known as quartet resonances, applies to gravity waves (including deep water). Case $N=5$ can also be relevant, as shown by \cite{dyachenko1994free}. Once solutions to the exact resonance equations are found, the dynamics of the wave amplitudes (at dominant orders) is dictated by coupled nonlinear PDE systems. In some simple scenarios the equations are integrable and can be solved analytically via the inverse scattering transform, leading to recurrent behaviour, showing periodic exchanges of energy between modes. In some scenarios, however, particularly in cases where multiple resonant interactions are coupled, the ODE and PDE systems obtained are known to display chaotic behaviour {\citep{1982PhFl...25.2159C,holmes1986chaotic,trillo1992modulational,kim1997chaotic}.}
 
For practical reasons, the multiple resonant interaction dynamics is discussed using stochastic approaches, quite successfully. Pioneered by \cite{Hasselmann:1962bn,Hasselmann:1963hua,Hasselmann:1963hu} and Zakharov (\cite{Zakharov:1967kn,zakharov1968stability}), the now established theory of wave turbulence describes the energy exchanges across scales due to resonant, {and quasi-resonant \citep{janssen2003nonlinear,annenkov2006role},} quartet interactions in the case of deep water surface gravity waves. A key assumption of this theory, in addition to the smallness of steepness and the limit of large box size, is an asymptotic closure of the hierarchy of cumulants \citep{newell2011wave}, which leads to a set of evolution equations for the so-called spectrum variables, namely the individual quadratic energies of the spatial Fourier transforms of the original field variables. This approach implies that the phases of the Fourier transforms do not play an important role in the dynamics of energy transfers, a scenario from which we wish to depart in this paper by investigating the evolution of the phases and how they interact with the spectrum variables.

Up to and including the papers by \cite{Perlin:1990gq}, \cite{Perlin:1991ee} and \cite{Perlin:1992jc}, and the review by \cite{HAMMACK:1993ht}, only two experiments studied resonant interactions between wavetrains with comparable initial amplitudes and only one experiment studied the relative roles of nonlinearity and randomness for a broad spectrum of waves. Since then, experiments on triad resonances in gravity-capillary waves were performed by \cite{Haudin:2016fq,2015PhRvL.114n4501A}. In the case of gravity waves, experiments on quartet resonances based on earlier work on modulational instability \citep{1967JFM....27..417B} were conducted by \cite{TULIN:1999il,SHEMER:1999fb}. Resonant interactions in the presence of an underwater current were investigated by \cite{Waseda:2015ft}. Experiments on persistent wave patterns and steady-state resonant waves were performed by  \cite{hammack2005progressive} and \cite{liu2015existence}. Finally, experiments on degenerate quartets of oblique waves not influenced by modulational instability were conducted by \cite{2016arXiv160609009B}. All these experiments have confirmed the resonant interaction theory in the case of small steepness ($k a < 0.1$ roughly). The latter experiments also confirmed the expected nonlinear corrections due to marginally off-resonance terms.

Our proposed numerical experiments and analyses differ from the above experiments and theories in three aspects: 

\begin{enumerate}
\item We obtain a strong nonlinear energy transfer mechanism for deep-water gravity surface waves, based on triad interactions, \emph{not} quartets.
\item A crucial concept is the \emph{precession frequency} of a triad, defined as the precession frequency, in the complex plane, of a certain product of the three modes involved. In other words, the precession frequency of a triad is a suitable linear combination of the precession frequencies, in the complex plane, of the individual modes' phases. These precession frequencies are generally nonlinear functions of the modes' spectrum variables and phases (see Section \ref{sec:precession_resonance} for a mathematical definition of precession resonance in the context of a $5$-mode Galerkin truncation). 
\item We demonstrate that the phase evolution affects the energy transfers across triads in a subtle but efficient way, via what we call \emph{precession resonance} \citep{Bustamante:2014jf}, a new resonance mechanism between two key frequencies: the frequency of oscillation of the spectrum variables and the triad precession frequency. Both frequencies are nonlinear and depend on the spectrum variables and phases.
\end{enumerate}

\vspace{2mm}

Our analysis has some features in common with previous theoretical work on the subject of nonlinear interactions: as will be shown in detail in subsequent sections, the precession frequency is obtained as a sum of a constant term plus an amplitude-dependent term. The constant term is merely the frequency detuning of the given triad. The amplitude-dependent term can be interpreted as a nonlinear correction to the frequency detuning, although its form does not reduce to the well-known frequency normalisation terms used in wave turbulence \citep{Newell:2001ue} or in strongly nonlinear models of wave turbulence \citep{lee2009renormalized}. 

There is, however, a significant difference. For surface gravity waves in deep water, we construct a resonance based on triads alone, which are known to have non-zero detuning. As nonlinear corrections are very small (due to the smallness of steepness) the precession frequency is also non-zero. Thus, a resonance is found when this precession frequency matches an oscillation frequency of the spectrum variables. In practice, we search for the precession resonance by exploring different choices for the wave-vectors until the resonance is found.

In this paper we consider a simplified setting of one-dimensional propagation of deep water surface gravity waves. While precession resonance is likely to be found in other settings such as two-dimensional propagation, capillary-gravity waves, finite depth, etc., we choose deep water and pure gravity waves because in that setting it is traditionally believed that triads do not play any significant role in resonant energy transfers. Additionally one-dimensional propagation is a simpler setting for both numerical and physical experiments. The results can be divided into two groups. In the first group we consider plane-wave interactions where periodic boundary conditions are assumed along the propagation coordinate. We demonstrate precession resonance in a reduced model of five nonlinearly interacting plane waves, leading to as much as 40\% energy transfers to the target wave at the peak of the resonance. We predict analytically the values of wave-vectors at which precession resonance is found and we confirm the prediction numerically. We then extend the demonstration to the full direct numerical simulation using a resolution of 8192 modes, confirming the first result. In the second group of results we consider a numerical wave tank of 50 metres length where five wave trains (resembling the five plane waves in the first part) are prepared and generated from the ends of the tank, to encounter and interact in the middle. Despite the fact that the waves have little time to interact, the resonance is demonstrated again. This reduced interaction window means that lower energy transfer efficiencies are obtained, which nevertheless should be large enough to be measured in physical experiments.

\section{Deep-water surface waves in one-dimensional propagation}

This section starts with a review of the third-order Hamiltonian theory of deep-water surface wave propagation in one dimension in natural variables (Sec.~\ref{sec:water_waves}), considers the equations of motion in normal variables (Sec.~\ref{sec:normal}) and then discusses a scenario of plane-wave scattering between five modes that shows the precession resonance effect (Sec.~\ref{sec:scattering}). Readers who are already familiar with the normal variables representation of water wave equations could skip this section.

\subsection{Third-order Hamiltonian theory in natural variables}
\label{sec:water_waves}
We consider the evolution of the free surface of a water column (in one dimension) via the boundary conditions for the surface elevation $\zeta(x,t)$ and the velocity potential $\phi(x,t)$. A widely used form of these conditions in the study of nonlinear phenomena on water waves is derived by expanding the vertical velocity at the free surface to third order \citep{zakharov1968stability,1995JFM...295..381C,Choi:2004uv,Tian:2010ku,Tian:2008bm}, leading to the system
\begin{align}
\label{eq:EOMa}
\partial_t{\zeta} + \cl \phi + \partial_x (\zeta \,\partial_x \phi) + \cl(\zeta \, \cl \phi) + \half \partial_{xx}(\zeta^2\cl \phi) + \half \cl (\zeta^2\partial_{xx} \phi) + \cl (\zeta \,\cl [\zeta\, \cl \phi]) &= 0,\\
\label{eq:EOMb}
\partial_t{\phi} + g\, \zeta + \half [(\partial_x \phi)^2 - (\cl \phi)^2] - [ \cl(\zeta\, \cl \phi) + \zeta\, \partial_{xx}\phi ] \cl\phi &=0\,,
\end{align}
where $g$ is the acceleration of gravity and the linear operator $\cl$ is defined in Fourier space by:
$$\cl \hat{\phi}_k(t) =- k \tanh(k h) \,\hat{\phi}_k(t), \qquad \hat{\phi}_k(t) = \frac{1}{L} \int_{-L/2}^{L/2} \phi(x,t) {\rm e}^{- i k x}~dx,$$
and $h$ $(>0)$ is the water depth. From here we restrict our analysis to the deep-water limit, $h \to \infty$, leaving the finite-depth case for a subsequent work. In the deep-water limit the $\cl$-operator's action reduces to
$$\cl \hat{\phi}_k(t) = - |k| \,\hat{\phi}_k(t).$$

Computationally we will assume periodic boundary conditions in the longitudinal direction with computational box size $L$, so that 
$$\phi(x+L,t) = \phi(x,t), \qquad \zeta(x+L,t) = \zeta(x,t), \qquad {\text{for all }} x, t \in \mathbb{R}.$$

Energy conservation is ensured by this system of equations, where energy is defined as
\begin{align}
H &= \frac{1}{L}\int_{-L/2}^{L/2} \mathcal{H}(x,t)~dx\,, \\
\mathcal{H}(x,t) & = \frac{ g \, \zeta^2}{2} - \frac{\phi}{2} \biggl[ \cl \phi + \partial_x (\zeta \,\partial_x \phi) + \cl(\zeta \, \cl \phi)  \nonumber \\   &  \qquad + \half \partial_{xx}(\zeta^2\cl \phi) + \half \cl (\zeta^2\partial_{xx} \phi) + \cl (\zeta \,\cl [\zeta\, \cl \phi]) \biggr ]  \,.
\end{align}
In fact, this energy serves as the Hamiltonian of the canonical formulation, best seen in Fourier space. Defining the Fourier components of surface elevation and velocity potential as
\begin{align}
\zeta(x,t) = \sum_{k \,\in \,(\frac{2\,\pi}{L})\mathbb{Z}} {\rm e}^{i k x} \hat{\zeta}_{k}(t)\,, \quad & \quad
\phi(x,t) = \sum_{k \,\in \,(\frac{2\,\pi}{L})\mathbb{Z}} {\rm e}^{i k x} \hat{\phi}_{k}(t)\,,
\end{align}
the system (\ref{eq:EOMa})--(\ref{eq:EOMb}) is equivalent to 
\begin{align}
\label{eq:H_eom}
\frac{\d}{\d t} \hat{\zeta}_{k}  = \frac{\partial H}{\partial \hat{\phi}_{k}^*} \,,\quad & \quad
\frac{\d}{\d t} \hat{\phi}_{k}  = - \frac{\partial H}{\partial \hat{\zeta}_{k}^*}\,,\quad
k \in \,\left(\frac{2\,\pi}{L}\right)\mathbb{Z}\,,
\end{align}
where $*$ denotes complex conjugation. Reality of the fields $\phi(x,t)$ and $\zeta(x,t)$ imply the identities
\begin{align}
\label{eq:real_zetphi}
\hat{\zeta}_{-k}  = \hat{\zeta}_{k}^* \,,\quad & \quad
\hat{\phi}_{-k}  = \hat{\phi}_{k}^*\,.
\end{align}

The explicit form of the Hamiltonian as sums of products of the Fourier components goes as follows:
$$H=H_2 + H_3 + H_4,$$
where $H_j$ is a homogeneous polynomial of degree $j$ in the Fourier components, defined by:
\begin{align}
\label{eq:H2}
H_2 &= \frac{1}{2}\sum_k  g |\hat{\zeta}_{k}|^2 + |k| |\hat{\phi}_{k}|^2\,,\\
\label{eq:H3}
H_3 &= -\frac{1}{2}\sum_{k_1,k_2}  (|{k_1}| \,|{k_1+k_2}| - k_1\,(k_1+k_2)) \hat{\phi}_{k_1} \hat{\zeta}_{k_2} \hat{\phi}_{k_1+k_2}^*\,,\\
\label{eq:H4}
H_4 &= -\frac{1}{4}\sum_{k_1,k_2,k_3}  |{k_1+k_2+k_3}||{k_1}|(|k_1|+|k_1+k_2+k_3| - 2 |{k_1+k_2}|) \hat{\phi}_{k_1} \hat{\zeta}_{k_2}\hat{\zeta}_{k_3} \hat{\phi}_{k_1+k_2+k_3}^*\,.\end{align}
The reality of the Hamiltonian is checked by noticing that the domains of the summations above are symmetric under reflections $k \to -k$.

\subsection{Galerkin truncations and equations of motion in normal variables}
\label{sec:normal}

We consider now the canonical transformation from natural variables $(\hat{\zeta}_{k}, \hat{\phi}_{k})$ to normal variables $a_k$, determined by the requirements that the quadratic part of the Hamiltonian be diagonal in the normal variables, and that the Poisson bracket be preserved. We have:
\begin{align}
\label{eq:normal1}
\hat{\zeta}_{k} &= \left({\frac{|k|}{4\, g}}\right)^{1/4}(a_k + a_{-k}^*)\,,\\
\label{eq:normal2}
\hat{\phi}_{k} &= - i \left({\frac{g}{4\, |k|}}\right)^{1/4}(a_k - a_{-k}^*)\,.
\end{align}
Notice that while $\hat{\zeta}_{k}$ and $\hat{\phi}_{k}$ satisfy the reality conditions (\ref{eq:real_zetphi}), the normal variables $a_k$ do not. In fact, we may interpret $a_k$ and $a_{-k}$ as independent waves propagating in direction $k$ and $-k$, respectively. Thus, the number of degrees of freedom is preserved under this linear transformation. {The variables $a_k$ have dimensions $L^{{3}/{2}}T^{-{1}/{2}}$.} 

We define a Galerkin truncation as a finite subset of  modes, namely a set ${\mathcal{C}} = \{K_1, -K_1, \ldots, K_N, -K_N\}$ ($N \in \mathbb{N}$) of wave-vectors along with their complex amplitudes $\hat{\zeta}_{K_j}$ and $\hat{\phi}_{K_j},$ which evolve according to the Hamiltonian equations (\ref{eq:H_eom}) with Hamiltonian given by equations (\ref{eq:H2})--(\ref{eq:H4}) but with the restriction that all wave-vectors involved belong to ${\mathcal{C}}.$ 

We remark that the set ${\mathcal{C}}$ {(called a \emph{cluster})} is made of pairs of wave-vectors related by parity transformation $k \to -k$, but one should not double-count: the fields $\hat{\zeta}_{-k}, \hat{\phi}_{-k}$ are not independent from the fields $\hat{\zeta}_{k}, \hat{\phi}_{k}$, however, the fields $a_k$ and $a_{-k}$ are independent. From here all sums over wave-vectors are restricted to the set ${\mathcal{C}}$.

We write the Hamiltonian in terms of the normal variables:
\begin{align}
\label{eq:H2_normal}
H_2 = &\sum_k  \omega_k a_k a_k^*\,,\\
\label{eq:H3_normal}
H_3 = &-\frac{1}{2}\sum_{k_1,k_2,k_3} \delta_{k_1 + k_2 - k_3}  \, V_{k_1 k_2 k_3} \, \left(a_{k_1} - a_{-k_1}^*\right) \, \left(a_{k_2} + a_{-k_2}^*\right) \, \left(a_{k_3}^* - a_{-k_3}\right)\,,\\
\label{eq:H4_normal}
H_4 =& -\frac{1}{4}\sum_{k_1,k_2,k_3, k_4} \delta_{k_1 + k_2 + k_3 - k_4}  \, V_{k_1 k_2 k_3 k_4} \,\nonumber \\ & \hspace{25mm}  \left(a_{k_1} - a_{-k_1}^*\right) \, \left(a_{k_2} + a_{-k_2}^*\right) \,\left(a_{k_3} + a_{-k_3}^*\right) \, \left(a_{k_4}^* - a_{-k_4}\right) \,,
\end{align}
where $\delta$ denotes the Kronecker delta, and the dispersion relation and interaction coefficients are defined as
\begin{align}
\label{eq:omega}
\omega_k &\equiv \sqrt{g\,|k|}\,,\\
\label{eq:V3}
V_{k_1 k_2 k_3} &\equiv  \frac{1 - \mathrm{sign}(k_1 k_3)}{2\sqrt{2}} g^{1/4}{\left|k_1^3{k_2}{{k_3^3}}\right|}^{1/4} \,,\\
\label{eq:V4}
V_{k_1 k_2 k_3 k_4} &\equiv ({|k_1|+|k_4|-|k_1+k_2|-|k_1+k_3|})\frac{{\left|{k_1^3{k_2 k_3 k_4^3}}\right|}^{1/4}}{4} \,.
\end{align}
These coefficients satisfy the following symmetries:
\begin{align}
\nonumber
V_{k_1 k_2 k_3} &= V_{k_3 k_2 k_1} = V_{-k_1, \,k_2, \,-k_3} = V_{k_1, \,-k_2, \,k_3}\,,\\
\nonumber
V_{k_1 k_2 k_3 k_4} &= V_{k_1 k_3 k_2 k_4} = V_{-k_1, -k_3, -k_2, k_4} = V_{k_1, k_3, k_2, -k_4}\,,
\end{align}
and the equations of motion read, after the transformation to normal variables:
\begin{align}
\label{eq:H_eom_normal}
i\,\frac{\d}{\d t}{a}_{k}  &= \frac{\partial H}{\partial a_{k}^*} = \omega_k a_k +   \frac{\partial H_3}{\partial a_{k}^*} +  \frac{\partial H_4}{\partial a_{k}^*} \,.
\end{align}
The explicit form of this right-hand side will be discussed in the next section.

\section{Deep-water plane-wave scattering scenario exhibiting precession resonance}
\label{sec:scattering}
In deep water waves nonlinearity is relatively weak mainly because of the tendency of the waves to break when their steepness is large. Historically this has been used as a justification for treating the nonlinear terms in a multi-scale formalism (the so-called weak nonlinearity limit). We will refrain from applying this formalism because this would entail the ``elimination'', via a near-identity transformation, of all terms in $H_3$ (the so-called triad interactions), including those terms that are relevant in the study of the precession resonance mechanism. 

{In order to describe the mechanism of precession resonance it is necessary to look at the amplitude-phase representation, where the phases $\phi_k \equiv \arg a_k$ play a significant dynamical role. These phases generally precess in time (in the complex plane) and, as we will see below, their precession frequencies interact with other frequencies in the system, such as the frequencies of the fluctuations of the spectrum variables $n_k \equiv |a_k|^2$. This interaction can lead to resonances when the system's parameters are chosen appropriately.}
For the present analysis of precession resonance (not for the simulations of the PDE which follows) we can discard the quartic part of the Hamiltonian $H_4$: its contribution is negligible because the wave steepness is small. The equations of motion (\ref{eq:H_eom_normal}) thus become:  
\begin{align}
\nonumber
 i\,\frac{\d}{\d t}{a}_{k}  =  \omega_k a_k +  \frac{1}{2}\sum_{k_1,k_2} \delta(k_1 + k_2 - k) &\big[  V_{k_1 k_2 k} \,  \left(a_{k_2} + a_{-k_2}^*\right) \, \left(a_{-k_1}^* - a_{k_1}\right) \\
\nonumber
   & +  V_{k_2 k_1 k} \, \left(a_{k_1} + a_{-k_1}^*\right) \, \left(a_{-k_2}^* - a_{k_2}\right)\\
\label{eq:H_3_system}
   & +  V_{- k_1 k k_2} \, \left(a_{-k_1}^* - a_{k_1} \right) \, \left(a_{-k_2}^* - a_{k_2}\right) \big]\,,
\end{align}
where
\begin{align}
\label{eq:L_deep}
V_{k_1 k_2 k_3} &\equiv  \frac{1 - \mathrm{sign}(k_1 k_3)}{2\sqrt{2}} g^{1/4}{\left|k_1^3{k_2}{{k_3^3}}\right|}^{1/4} \,,
\end{align}
and we remark that this coefficient is non-zero only if $k_1$ and $k_3$ have opposite signs.

\subsection{A $5$-mode Galerkin truncation} 
\label{sec:5mode}
We discover, empirically, that the smallest system able to exhibit the precession resonance is a Galerkin truncation consisting of only {five} interacting plane waves, with wavenumbers
$$K_1, K_2, K_3, -K_1, -K_2, \qquad \text{where} \quad K_3 = K_1 + K_2,$$
and $K_1$ and $K_2$ are positive but independent. The reason for this choice will be explained in due course.

Let us denote the five complex amplitudes corresponding to these five modes as follows:
$$a_j \equiv a_{K_j},  \quad j=1,2,3 \qquad a_{-j} \equiv a_{-K_j}, \quad j=1,2\,.$$ 
From equations (\ref{eq:H_3_system})--(\ref{eq:L_deep}), these amplitudes satisfy a coupled system of ordinary differential equations:
\begin{align}
\label{eq:8-mode_1}
 i\,\frac{\d}{\d t}{a}_{1}  &=  \omega_{1} a_{1} +   V\, a_{3} \, \left(a_{2}^* - a_{-2}\right)\,, \quad      i\,\frac{\d}{\d t}{a}_{-1}  =  \omega_{1} a_{-1} +   V\, a_{3}^* \, \left(a_{-2}^* - a_{2}\right)\,,\\
\label{eq:8-mode_2}
   i\,\frac{\d}{\d t}{a}_{2}  &=  \omega_{2} a_{2} +  V \, a_{3} \, \left(a_{1}^* - a_{-1}\right)\,, \quad    i\,\frac{\d}{\d t}{a}_{-2}  =  \omega_{2} a_{-2} +  V \,a_{3}^* \, \left(a_{-1}^* - a_{1}\right)\,,\\
\label{eq:8-mode_3}
   i\,\frac{\d}{\d t}{a}_{3}  &=  \omega_{3} a_{3} +   V \, \left(a_{-1}^* - a_{1} \right) \, \left(a_{-2}^* - a_{2}\right)\,, 
\end{align}
where $\omega_j \equiv \omega(K_j) = \sqrt{g K_j}\,, \,\, j=1,2,3,$ and 
$$V = V_{-K_1 K_3 K_2} =\frac{1}{\sqrt{2}} {\left(g{K_1^3{K_2}^3}  {K_3}\right)}^{1/4}\,.$$

\begin{figure}
\begin{center}
\includegraphics[width=0.4\textwidth]{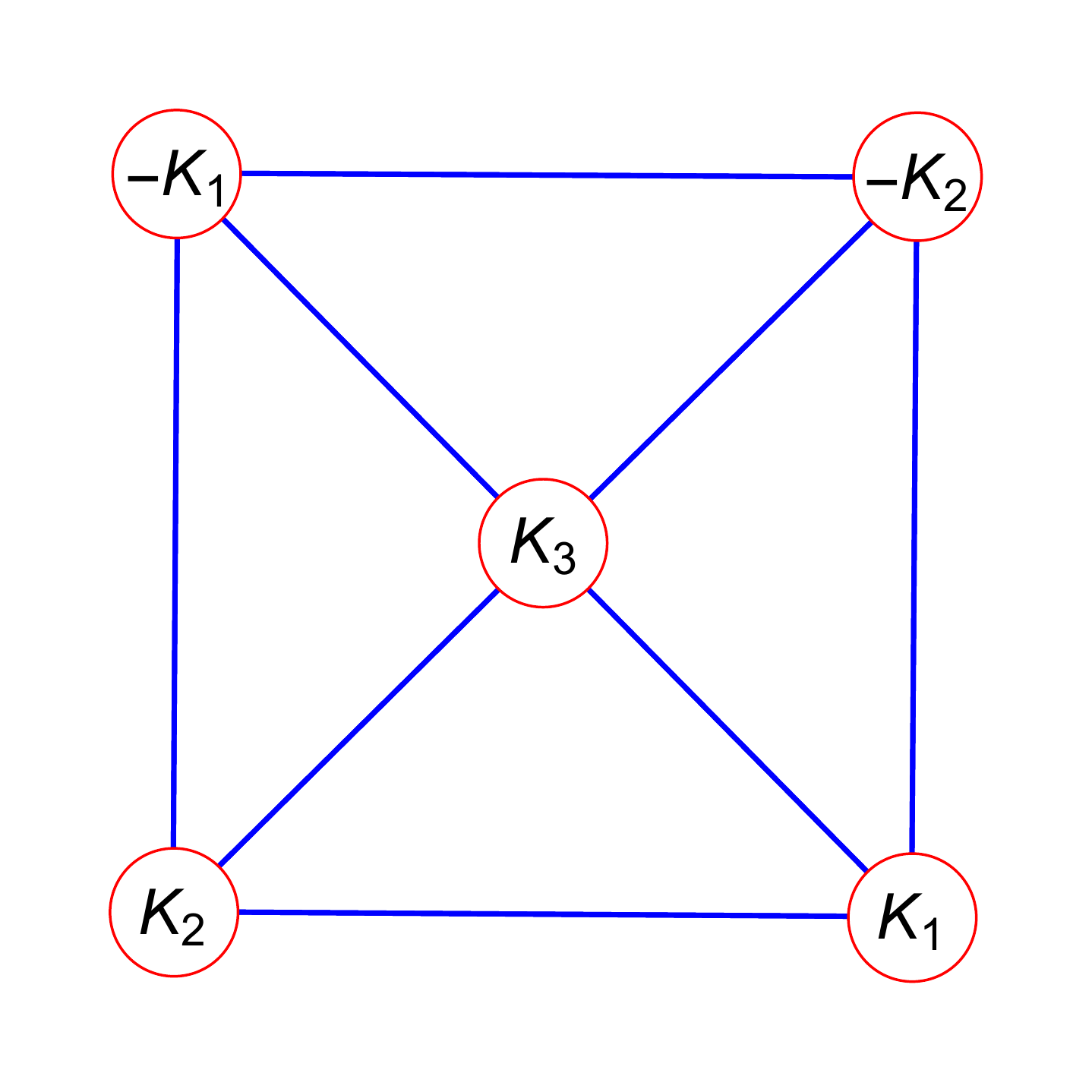}
\end{center}
\caption{\label{fig:5-mode} Graphical representation of the $5$-mode system under consideration. Nodes of the graph represent the modes and the vertices their connections, i.e. smallest triangles represent the four triad interactions.}
\end{figure}

Figure \ref{fig:5-mode} illustrates the usual picture of triad interactions stemming from the nonlinear terms in system (\ref{eq:8-mode_1})--(\ref{eq:8-mode_3}). This consists of four triads, denoted by the four triplets of wave-vectors $(\pm K_1, \pm K_2,  K_3)$ and graphed as the four small triangles in the figure. There is an intricate structure of connections between modes and triads: mode $a_3$ belongs to all triads, and the other four modes belong to two triads each. For example, mode $a_{-{1}}$ belongs to the triads $(-K_{1},\pm K_2,  K_3).$  Each \emph{triad} is connected to two triads via two common mode connections, and to {one triad} via one common mode ($K_3$) connection.

This structure provides a rich nonlinear behaviour which is not integrable, unlike the case of the isolated triad. To describe the precession resonance theory {\citep{Bustamante:2014jf}} we work on the amplitude-phase representation and consider the number of degrees of freedom of this system:
$$a_j = |a_j| \exp(i\,\phi_j)\,,\quad j = \pm1, \pm2, 3\,.$$
Rewriting evolution equations (\ref{eq:8-mode_1})--(\ref{eq:8-mode_3}) shows that the degrees of freedom are five real amplitudes:
$$|a_j|\,,\quad j = \pm1, \pm2, 3\,,$$
and three independent combinations of phases (so-called triad phases). We can choose the following basis for the phases:
$$\varphi_1 = \phi_3 - \phi_2 + \phi_{-1}\,, \quad \varphi_2 = \phi_3 - \phi_2 - \phi_1\,, \quad \varphi_3 = \phi_{3} + \phi_{-2} - \phi_1\,.$$
In addition, the system has two independent quadratic invariants, which can be obtained using the method introduced in \cite{Harper:2013gl},
$$I_1 = |a_1|^2 + |a_3|^2 - |a_{-1}|^2\,, \qquad I_2 = |a_2|^2 + |a_3|^2 - |a_{-2}|^2 \,,$$
and the Galerkin-truncated Hamiltonian is a cubic invariant:
\begin{equation}
\label{eq:H_3_6-mode}
H = \sum_{j=1}^3\omega_j |a_j|^2 + \sum_{j=1}^2\omega_j |a_{-j}|^2 + V \left[(a_{1}^* - a_{-1}) (a_{2}^* - a_{-2})a_{3} + \mathrm{c.c.}\right]\,.
\end{equation}
Therefore the system is effectively five dimensional (five real amplitudes, three triad phases, minus three invariants).

\subsection{Precession resonance}
\label{sec:precession_resonance}
Studying the behaviour of system (\ref{eq:8-mode_1})--(\ref{eq:8-mode_3}) as a function of the wavenumbers $K_1$ and $K_2$ ($K_3 = K_1 + K_2$ is not independent) reveals a precession resonance involving the triad $(-K_1,K_2,K_3)$, namely the modes $a_{-1}, a_2, a_3$  which interact through the term of the cubic Hamiltonian $V [a_{-1} a_2^*a_3 + \mathrm{c.c.}] = 2 V |a_{-1}| |a_2| |a_3| \cos(\varphi_1) = 2 V |a_{-1}| |a_2| |a_3| \cos(\phi_3 - \phi_2 + \phi_{-1})$.

Consider for simplicity energy transfers towards mode $a_{-1}$ due to the above triad interaction. The rate of change of this mode's energy is given by
\begin{equation}
\label{eq:energy(-1)}
\frac{d}{dt}  |a_{-1}|^2 = 2 V  |a_{-1}| |a_2| |a_3|  \sin (\phi_3 - \phi_2 + \phi_{-1}) + \mathrm{other \,\, triad \,\, interactions}\,.
\end{equation}
By definition, a precession resonance occurs when the time series of the amplitude term $|a_{-1}| |a_2| |a_3|$, in the right-hand side of equation (\ref{eq:energy(-1)}), has a salient frequency $\Gamma$ that coincides with the precession frequency of the phase  $\varphi_1$, namely the frequency
$$\Omega \equiv \frac{1}{T}\int_0^T \dot \varphi_1 dt = \frac{1}{T}\int_0^T (\dot\phi_3 - \dot\phi_2 + \dot\phi_{-1}) dt\,,$$ 
where $T$ is a sufficiently long integration time (typically larger than the characteristic nonlinear time scale $t_{NL} \equiv 2\pi/\Gamma$).

More specifically, let us assume the amplitude term has a leading time dependence of the form
$$|a_{-1}| |a_2| |a_3|  \approx \overline{|a_{-1}| |a_2| |a_3|} + R \sin(\Gamma t + \phi_0),$$
where $R (>0), \phi_0$ are real constants and over-bar denotes the time average. {Similarly, we can approximate the phase $\varphi_1$ by the expression $\varphi_1 \approx \Omega t + \phi_1$, where $\phi_1$ is a real constant.} A precession resonance is signalled when 
$$\Gamma = \Omega.$$
The effect of this resonance on the energy transfer towards $a_{-1}$ is seen by looking at the time dependence of the right-hand side of equation (\ref{eq:energy(-1)}), which gains a zero-mode. Discarding other triad interactions (which generically are off-resonance), we obtain:
\begin{eqnarray*}
\frac{d}{dt}  |a_{-1}|^2 &\approx& 2 V  \left[\overline{|a_{-1}| |a_2| |a_3|} + R \sin(\Gamma t + \phi_0)\right] \sin (\Omega t + \phi_1) + \mathrm{nonresonant \,\, terms}\\
&\approx& V R  \cos([\Gamma - \Omega] t + \phi_0-\phi_1) + \mathrm{nonresonant \,\, terms}\,.
\end{eqnarray*}
Near precession resonance $\Gamma - \Omega$ is very small and so the right-hand side will vary slowly, leading to a sustained linear growth (or decay, depending on the sign of $\cos (\phi_0-\phi_1)$) of the energy $|a_{-1}|^2$.  An analogous analysis shows that the energies of the other modes in the triad ($|a_{2}|^2$ and $|a_{3}|^2$) compensate this variation via a detailed balanced exchange. 

The maximal time scale at which we could possibly expect to see a sustained growth is given approximately by
\begin{equation}
\label{eq:Tmax}
T_{\max} = \frac{2 \pi}{|\Gamma - \Omega|}\,.
\end{equation}
Of course, this linear energy growth/decay will eventually lead to a recurrent nonlinear oscillation so the practical time scale of the resonance should be smaller than $T_{\max}$.

The final aspect to consider is how to locate such a resonance in a space of initial conditions spanning choices for not only the $a_j$\,s but also the $K_j$\,s themselves.  Nonlinear frequencies in the time series of $|a_{-1}| |a_2| |a_3|$ will appear as combinations of the linear frequencies $\omega_j$ generated by the nonlinear interactions, with some corrections that are small at low amplitudes. In particular we find that 
$$\Gamma \approx 2\omega_2 = 2 \sqrt{g} \sqrt{K_2}\,$$
is the salient frequency involved in the example of precession resonance that we will investigate.\\

We find precession frequencies, $\Omega$, have a dominant contribution from the linear frequency mismatch at low amplitudes. We obtain, to leading order,
$$\Omega \approx \omega_3 - \omega_2 + \omega_1 = \sqrt{g} \left(\sqrt{K_1+K_2}-\sqrt{K_2}+\sqrt{K_1}\right).$$

The above approximations reduce the condition of precession resonance $\Gamma = \Omega$ at low amplitudes to the simple algebraic relation
\begin{align}
 \omega_3 - 3\omega_2 + \omega_1 = 0   \quad \Longrightarrow \quad \sqrt{K_1 + K_2} - 3 \sqrt{K_2} + \sqrt{K_1} = 0\,,\label{eq:res1} 
 \end{align}
whose solution is parameterised as 
$$K_1 = 16 p\,,\quad K_2 = 9 p\,, \quad p \in \left(\frac{2\pi}{L}\right)\mathbb{N}$$
{(notice that $\frac{2\pi}{L}$ is the wave-vector corresponding to the full computational box of length $L$).} This condition has been determined by noticing that many such simple relations may allow a resonance at low amplitude and experimenting numerically to determine where such values of $\Gamma$ are realised. Thus five modes are the minimal Galerkin truncation allowing for resonance; if any further modes are truncated we no longer observe $\Gamma$ in the frequency spectrum of $|a_{-1}| |a_2| |a_3|$ and consequently the resonance does not occur. 

\subsection{Numerical study}
\label{sec:5modeNUM}

Given appropriate initial conditions, we expect to observe strong transfers in any system containing the modes $a_{\pm1}, a_{\pm2}, a_3$ {when $\alpha$ $(\equiv K_1/K_2)$ is close to $\alpha_c$ $(\equiv 16/9)$.} We proceed henceforth with a numerical investigation, using a pseudospectral method to solve equations (\ref{eq:EOMa})--(\ref{eq:EOMb}) truncated to the above five modes. Note this is slightly different than system (\ref{eq:8-mode_1})--(\ref{eq:8-mode_3}) as the quartic terms from $H_4$ are retained. 

In order to probe this resonance we evolve this system for a range of values of the wave-vectors $K_1, K_2$ using a parameterisation which intersects the resonance curve $K_1/K_2 = 16/9$ orthogonally and at values of $K_j$ which are physically relevant to prospective experiments, namely 
$$K_2 = -\frac{9}{16}K_1 + K_0\,, \quad  630 \frac{2\pi}{L} \leq K_1 \leq 653 \frac{2\pi}{L} \,, \quad \mathrm{with} \quad K_0=720\frac{2\pi}{L} \,.$$
{with a domain $L=50 m$ (metres). We initialise the system with randomised phases and with amplitudes 
$$|a_{\pm 1}|=6\times10^{-9}A \,\left(\frac{m^{3}}{s}\right)^{1/2}, \,\quad |a_{\pm 2}|=5\times10^{-4}A\,\left(\frac{m^{3}}{s}\right)^{1/2}, \,\quad |a_{3}|=1.5\times10^{-5}A\,\left(\frac{m^{3}}{s}\right)^{1/2},$$
where $A$ is an overall amplitude scale parameter (dimensionless) and $s$ stands for seconds.} We evolve for 500 seconds in the first instance and examine modes $a_{-1}$ and $a_{3}$ separately for their contribution to the quadratic component of the total energy:

\beq
\label{eq:eff}
 \mathcal{E}_{K}(t) = \frac{\omega_K a_K a_K^*}{H}\,.
 \eeq 

\begin{figure}
\begin{center}
\scalebox{0.4}{\input{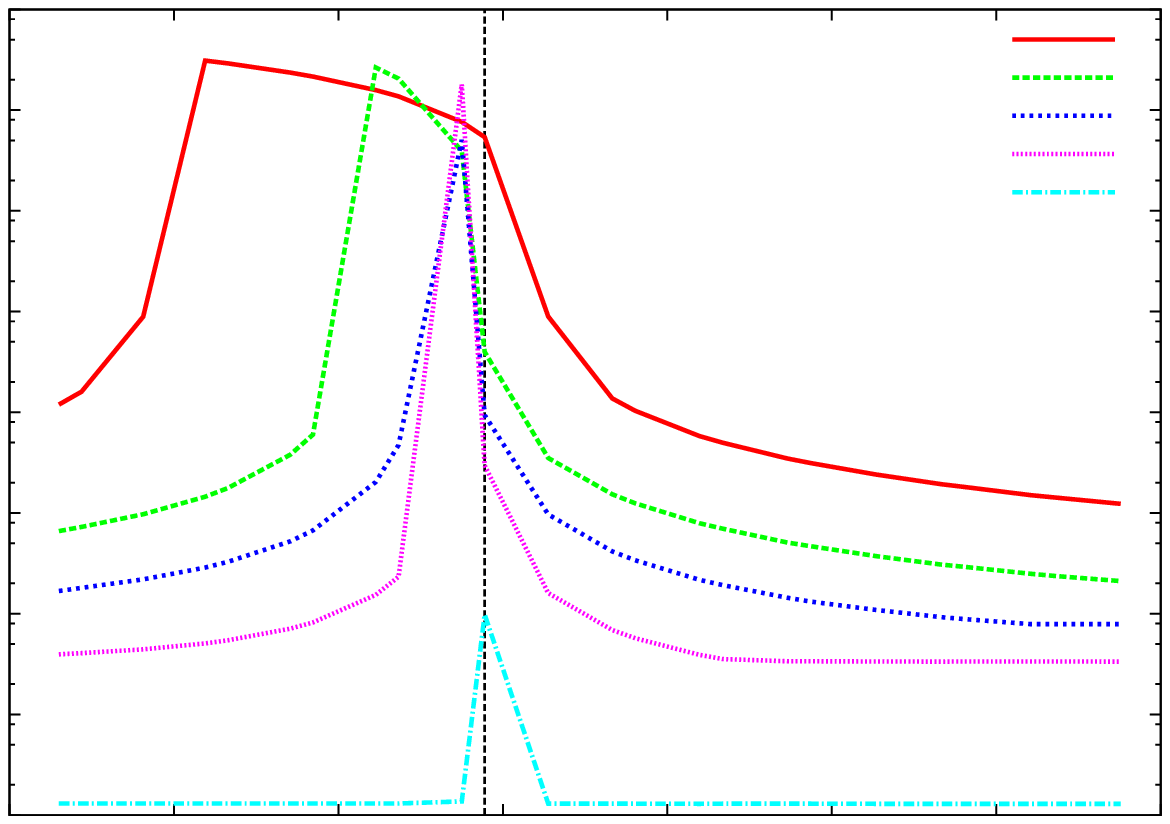}}
\scalebox{0.4}{\input{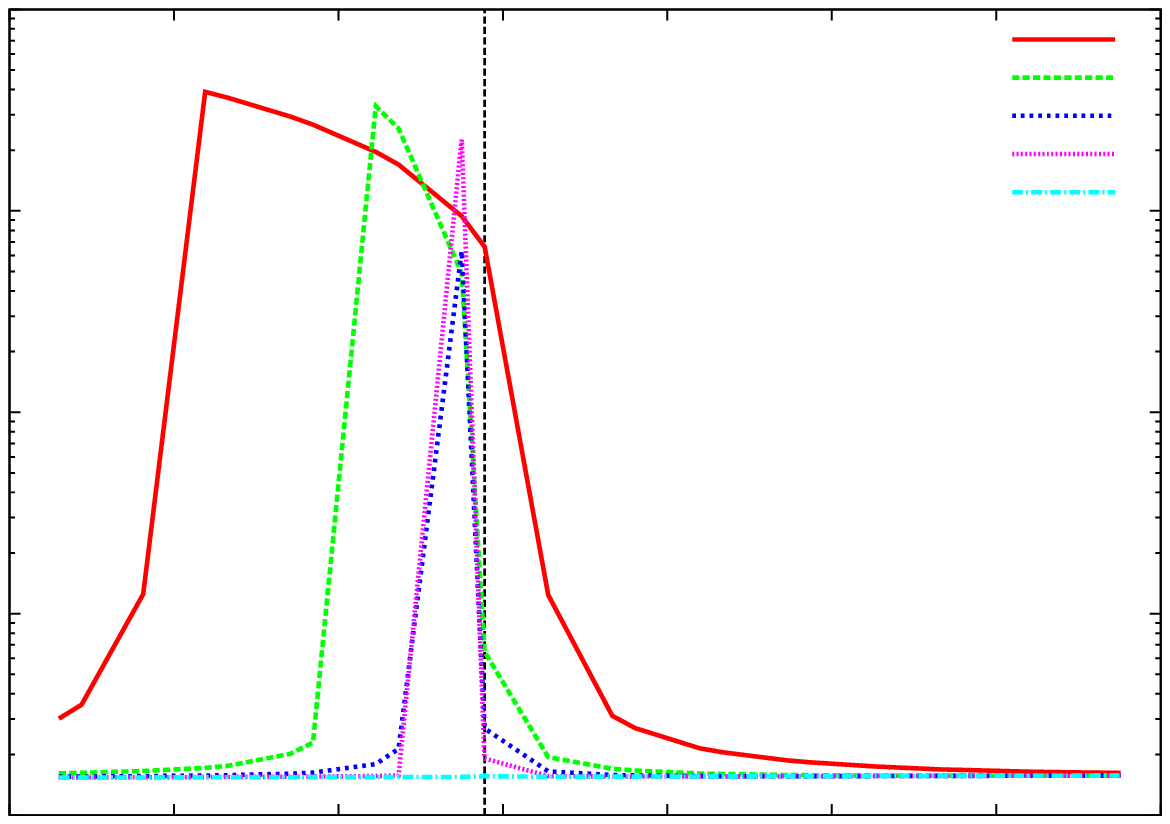}}\\
\vspace{-20mm}
\scalebox{0.4}{\input{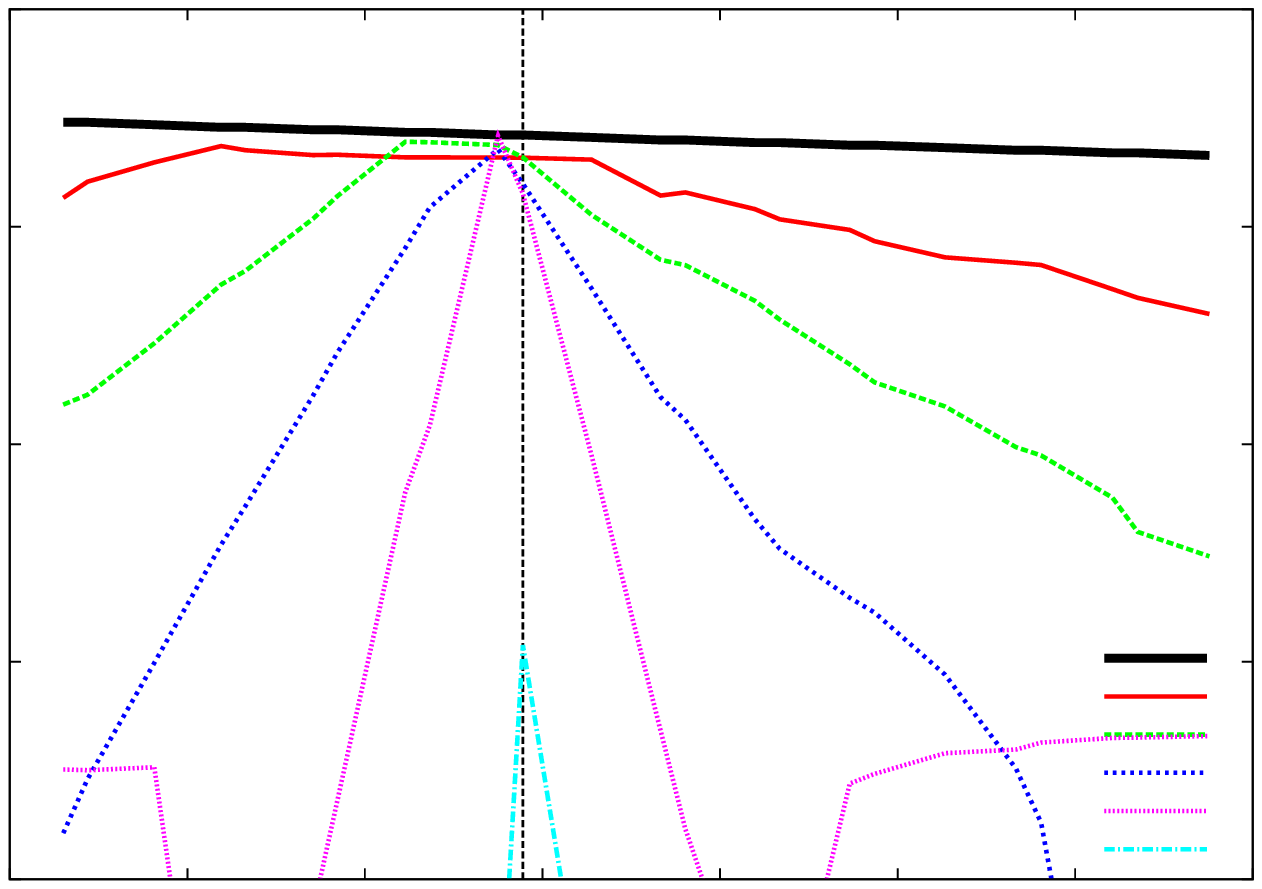}}
\scalebox{0.4}{\input{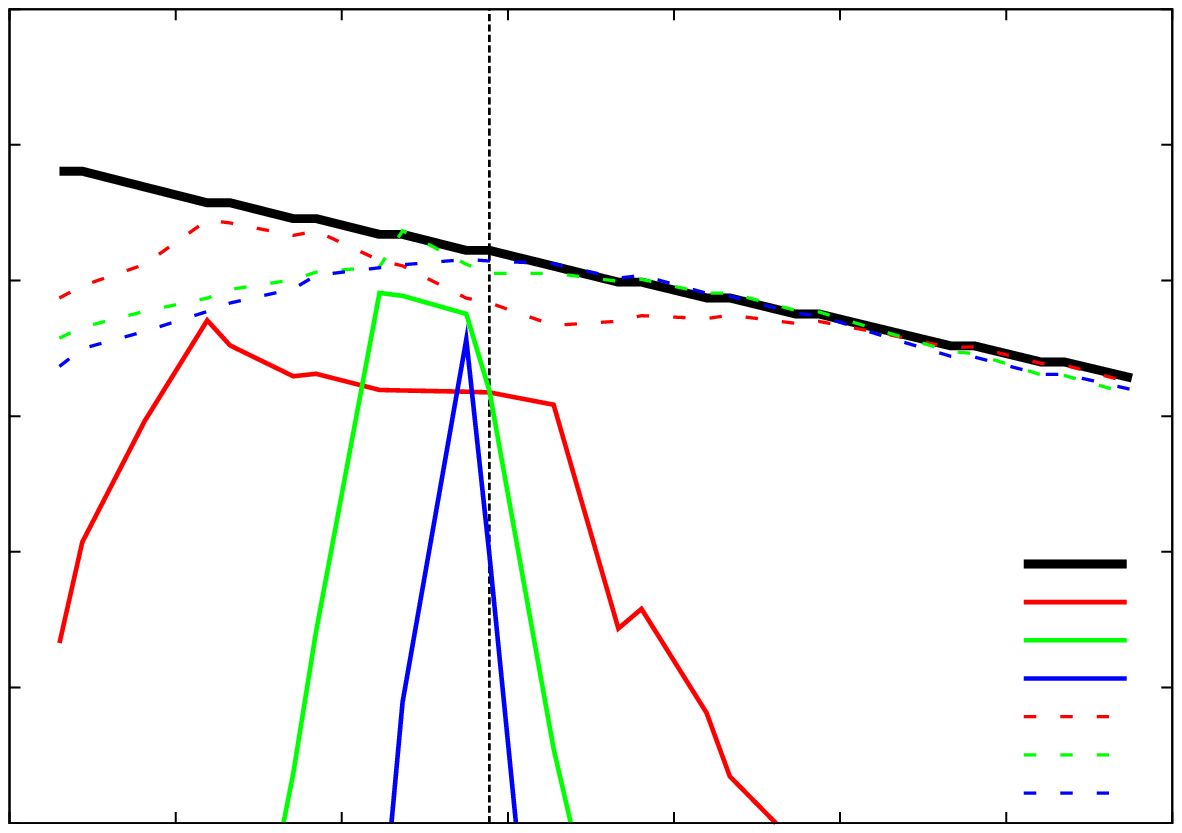}}
\end{center}
\vspace{-10mm}
\caption{ \textbf{Upper:} Efficiency {plots} for five values of initial condition rescaling: $A = 0.2, 1, 1.5, 2, 3$, showing the shift from the predicted resonance at $\alpha= K_1/K_2 =\alpha_c =16/9$ (vertical lines) as increased nonlinearity contributes to precession. \textbf{Lower Left:} Precession frequency $\Omega \equiv \langle \dot\phi_3 - \dot\phi_2 + \dot\phi_{-1}\rangle$ (time average over $500s$) for the same five values of $A$ along with the target frequency $2\omega_2$ showing the matching at the efficiency peaks, the broadening accounted for by the proximity to the target frequency over a range of $\alpha.$ \textbf{Lower Right:} Behaviour near the resonance for larger $A$, including an estimate of the nonlinear frequency $\Gamma$ as measured from the spectrum of product $|a_{-1}| |a_2| |a_3|.$ This shows the correction to both $\Omega$ and $\Gamma$ as $A$ increases. \label{5mode1}}
\end{figure}

Figure \ref{5mode1} shows the outcome of such sweeps in terms of the ratio 
$\alpha$,
plotting $\max_t \mathcal{E}_K$ for a range of initial amplitudes, $A$, and plotting the  precession frequency for each case. One observes that in the low amplitude limit, the resonance peak is smaller but coincides precisely with the predicted ratio $\alpha = \alpha_c (\approx 1.778)$. Similarly the precession frequency is observed to approach $\Gamma (\approx 2\omega_2)$ when there is a peak in the energy transfer. As the amplitude factor $A$ increases, the overall energy transfer grows significantly, {reaching as far as $40\%$ transfer efficiency,} and at the same time the peak is shifted slightly from $\alpha_c$ and broadens. This coincides with the precession frequency `hugging' the predicted target frequency for a larger range of $\alpha$ in the vicinity of $\alpha_c$. As the nonlinearity is increased, {both the precession and nonlinear frequencies} accumulate contributions from additional interactions which serve as ``corrections'' on top of the base frequency mismatch, thereby displacing and broadening the point of resonance.  Notice that the overall trend as $A$ increases is not monotonic; $A=1$ shows stronger transfer than $A=1.5.$ This can be explained by the discrete wavenumber/frequency space of the sweep in $\alpha$; nonlinear corrections move the precise point of resonance off the integer wavenumber grid. In the physical system a continuous frequency space will produce a continuous shift and broadening of the peak.

\begin{figure}
\scalebox{0.29}{\input{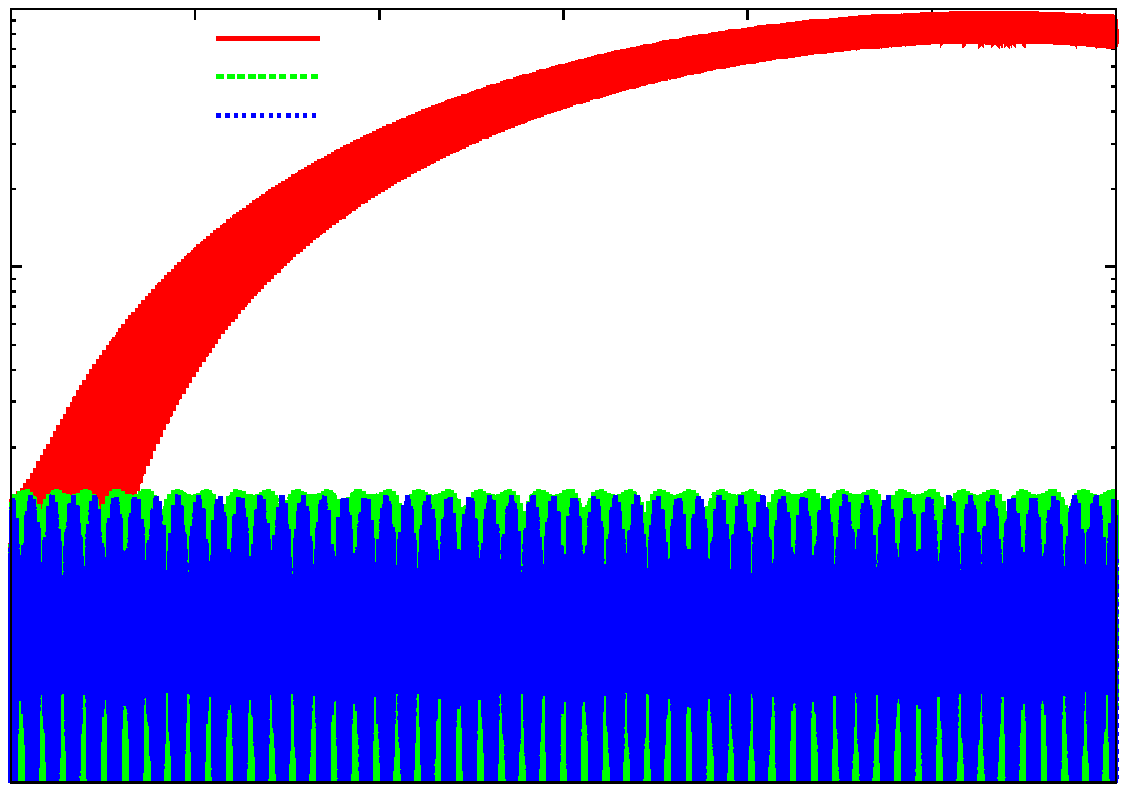}}
\scalebox{0.29}{\input{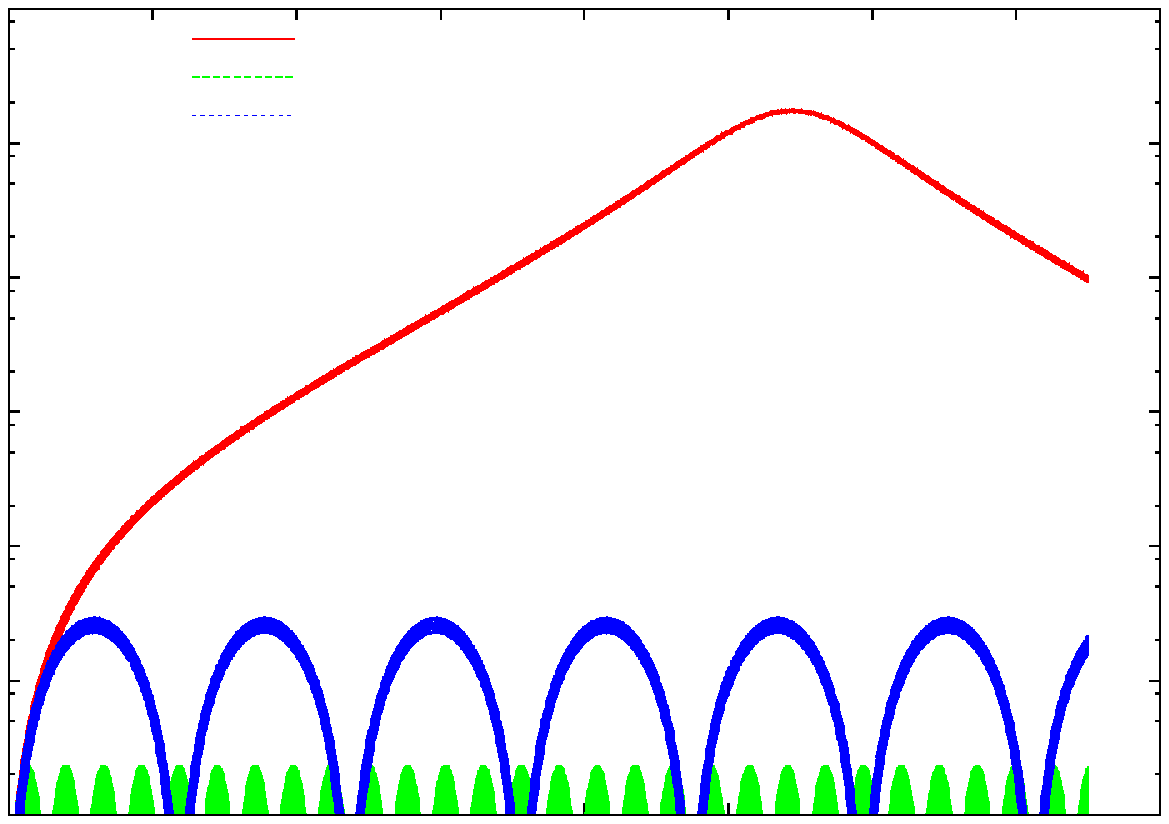}}
\scalebox{0.29}{\input{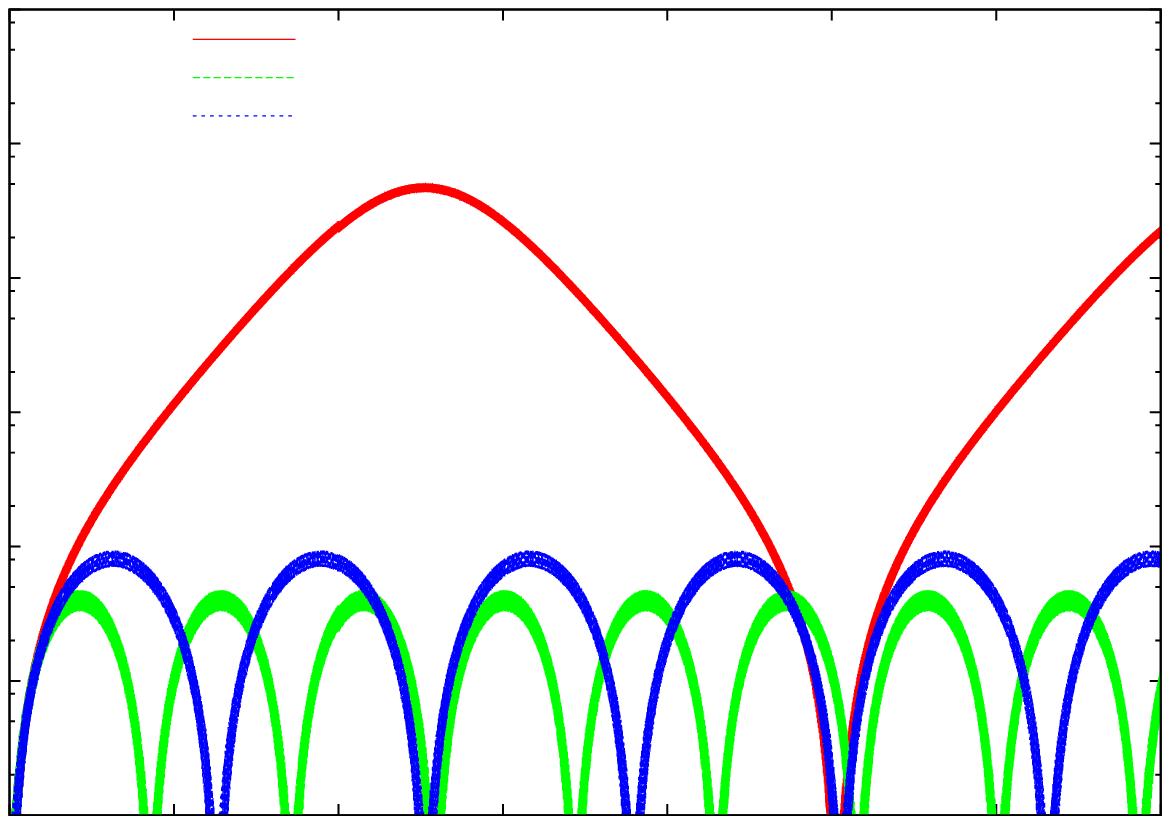}}
\caption{\label{5mode2}Time series of the efficiency ${\mathcal{E}}_{-K_1}(t)$ for $A=0.2$ (left), $A=1$ (centre) and $A=1.5$ (right) and each for three values of the ratio $\alpha \equiv K_1/K_2$, showing the resonant growth and two non-resonant examples either side. Note ${\mathcal{E}}_{-K_1}(t)$ is shown in logarithmic scale and the total time integration varies to account for the longer timescales at low amplitude.}
\end{figure}

To demonstrate the typical timescales involved, figure \ref{5mode2} shows time series of the efficiency $\mathcal{E}_{-K_1}$ for resonant and non-resonant values of $\alpha$ and for $A=0.2, \,\, 1$ and $1.5.$ As one might expect the growth rate is increased for the larger amplitude, higher nonlinearity case. We found that the lower growth rate at low $A$ meant that longer time integrations were required to reach the peak, recurrent, resonant behaviour. At $A=0.5$ this required $T=1500 s$ and at $A=0.1$, $T\approx3000 s$.

An argument could be put forward that the resonance condition (\ref{eq:res1}) appears in the weakly nonlinear limit as a fifth-order resonance (quintet) in the normal-form Hamiltonian, i.e. after all triads and non-resonant quartets are eliminated from the cubic Hamiltonian (\ref{eq:H_3_6-mode}) via a near-identity transformation, leading to a quintic Hamiltonian. If this was the explanation for the phenomenon observed here then the timescale of the resonant energy transfer would be proportional to the inverse third power of the amplitude rescaling. 
As seen in figure \ref{5mode2} the timescale is reduced by a factor of {three} (roughly) when the amplitude is {augmented} by a factor of {five}, confirming that the growth is proportional to the inverse power of the amplitude rescaling and so the dominant mechanism is a triad interaction. We remark that this demonstrates that it is not simply a normal-form quintet resonance but rather a triad precession resonance, {for if it was a quintet resonance, then the timescale would be reduced by a factor of over a hundred (and not a factor three as observed).}

\section{Precession resonance in the PDE}
\label{sec:PDE}
\subsection{Plane-wave scattering scenario in high resolution}
\begin{figure}
\begin{center}
\scalebox{0.33}{\input{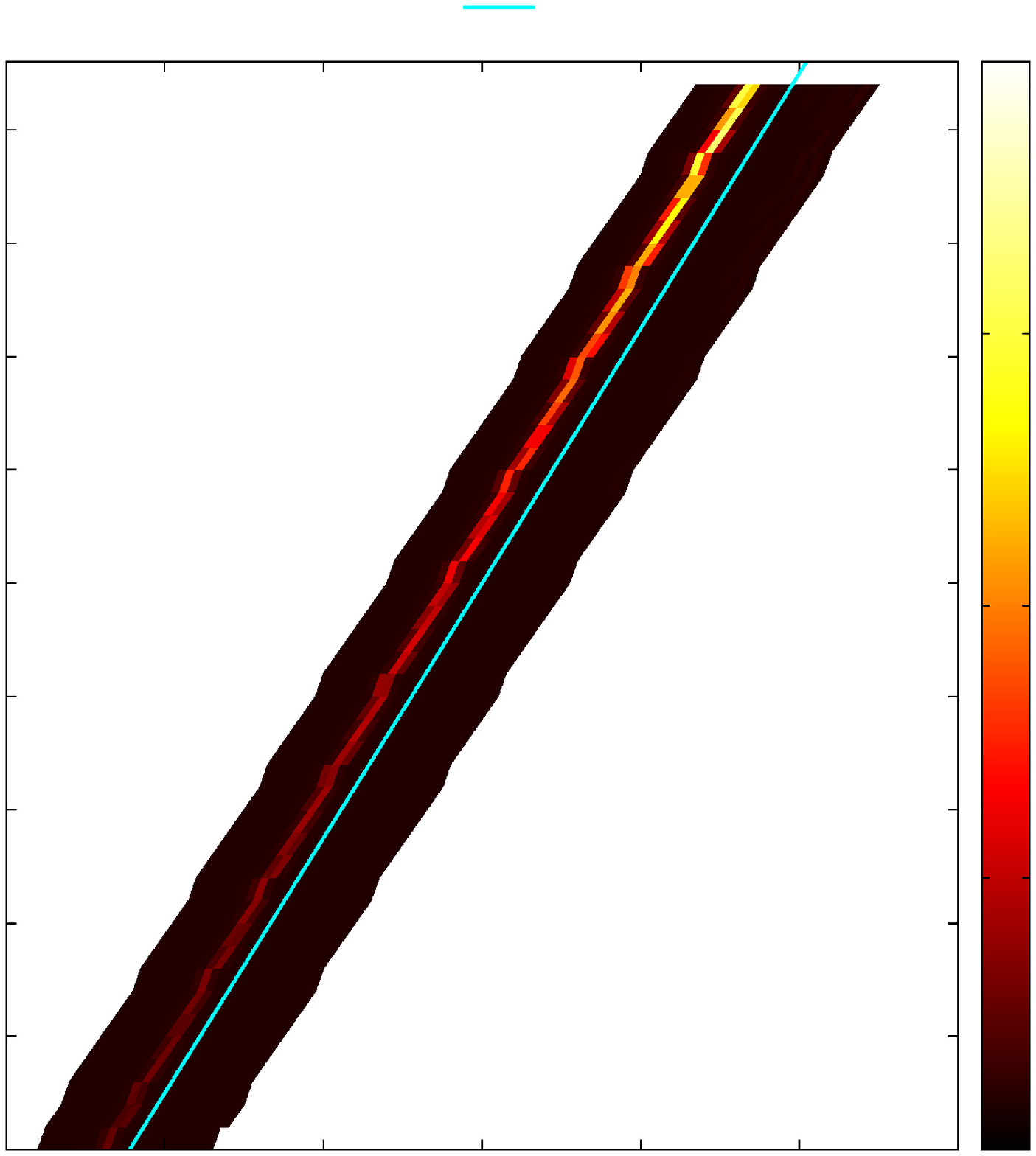}
\input{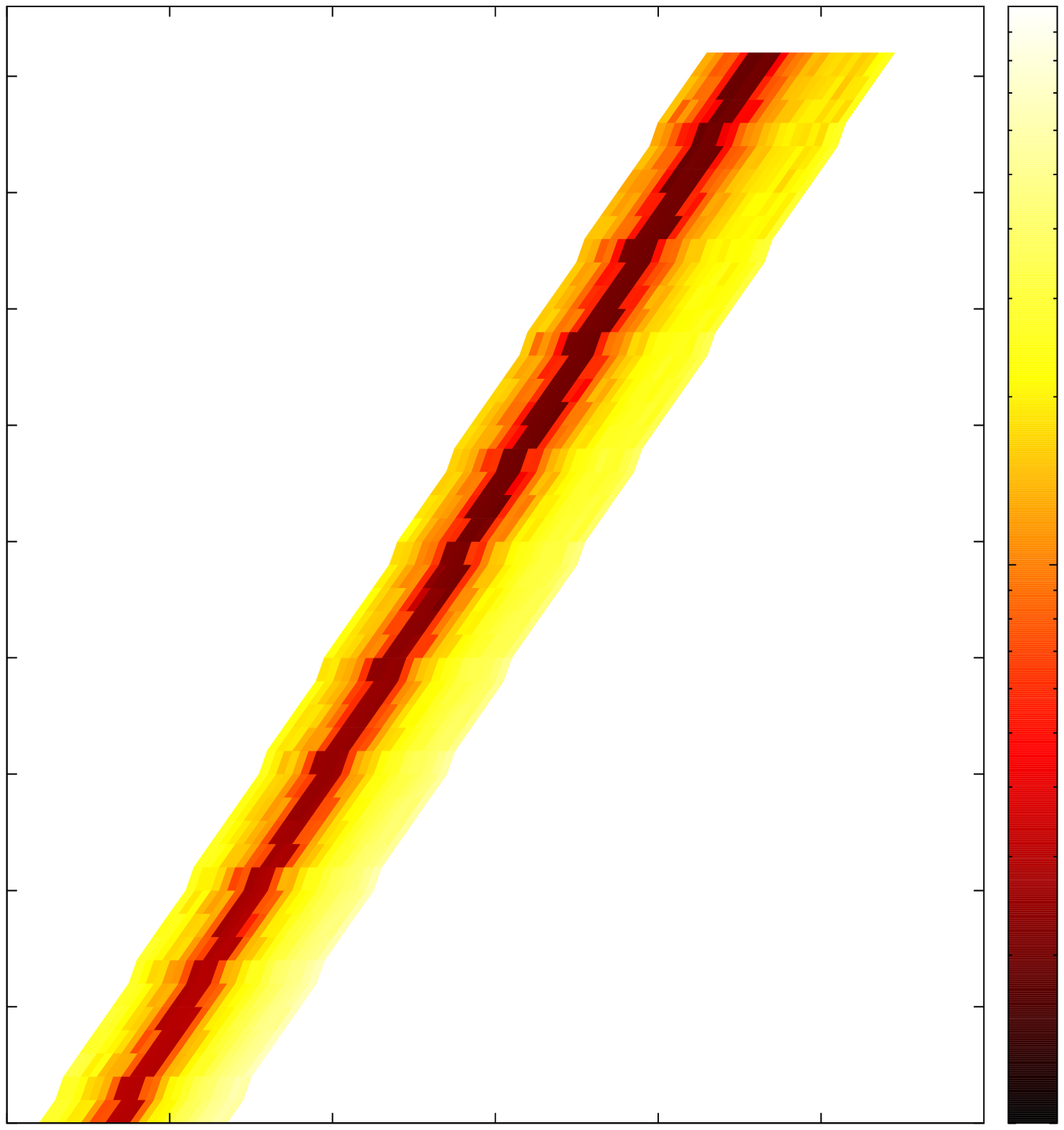}}
\end{center}
\caption{\label{fig:2D}Results from a sweep of $K_1$ and $K_2$ initial condition modes for the full PDE system. \textbf{Left:} maximum efficiency achieved over a simulation time of 200 seconds. \textbf{Right:} relative error between the computed precession frequency $\Omega$ and the target nonlinear frequency $\Gamma \approx 2\omega_2.$ The precession shows a transition to the target frequency precisely where the target mode exhibits increased transfer efficiency.}
\end{figure}

Having established the existence of precession resonance in a simplified reduced model, we now extend our numerical experiments to model the full PDE system {(\ref{eq:EOMa})--(\ref{eq:EOMb}). We retain the initial conditions from Section \ref{sec:5modeNUM}, i.e. energy on $\pm K_1,$ $\pm K_2$, and $ K_3$ but now every mode is free to evolve in a $N=8192$ fully-dealiased discretisation, formally analogous to system (\ref{eq:H2_normal})--(\ref{eq:H_eom_normal}) with cluster ${\mathcal C} = \{\pm K_j\}_{j=1}^{2730}$. We integrate for $T=200s$ and set $A=1.$ In order to demonstrate the robustness and persistence of the resonance across a range of relevant wavenumbers we show in figure \ref{fig:2D} a sweep across a strip on the $(K_1, K_2)$ plane (chosen to minimise computations but show the key effect) where colours indicate the energy transfer efficiency, i.e. $\max_t \mathcal{E}_{-K_1}(t).$ Over this we superimpose the predicted resonance curve $\frac{K_1}{K_2}=\alpha_c=\frac{16}{9}.$ Notice that, as in the high-nonlinearity $5$-mode case, the peak has shifted with respect to $\alpha_c$, to approximately $\frac{K_1}{K_2}=1.764$, i.e. by less than one percent.} This gives us confidence that this is the same mechanism outlined in the previous section. Furthermore figure \ref{fig:2D} also shows a relative error between precession frequency and target frequency across the same 2D strip. There is a clear sharp minimum at $\frac{K_1}{K_2}$ coincident with the resonant transfer increase. 

\subsection{Encountering wave packets: {high-resolution numerical wave tank}}
\begin{figure}
\begin{center}
\scalebox{0.75}{\input{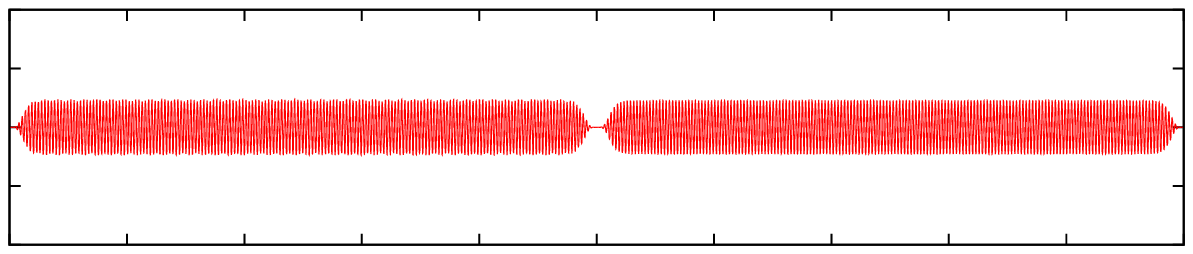}}\\
\vspace{-7mm}
\scalebox{0.75}{\input{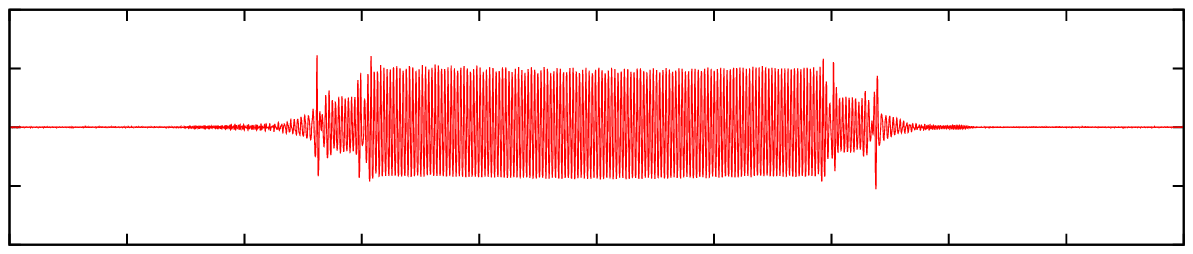}}\\
\vspace{-7mm}
\scalebox{0.75}{\input{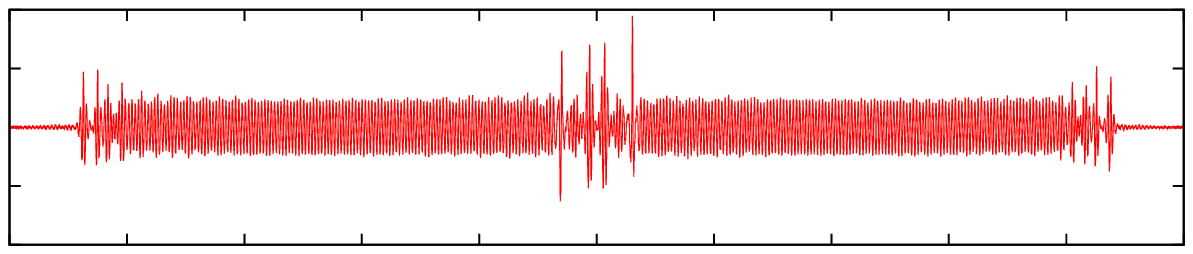}}\\
\end{center}
\caption{\label{fig:eta} Surface elevation $\zeta$ for $\alpha=1.754$ at $t=0, \,\, 50$ and $100$ from top to bottom.}
\end{figure}

In this final results section we investigate the numerical simulation of coincident packets of waves containing the initial frequencies under consideration. The motivation here is to provide experimentalists with evidence that this effect is physically observable, in particular in typical lab-based wave tank geometries  {(in fact, our preliminary experiments seem to confirm the precession resonance mechanism and we will present a detailed account of experimental results in a subsequent paper).} The wave packets are prepared by convolving a plane wave profile, containing the requisite modes, with a smoothing kernel which produces the necessary half domain localised left or right propagating waves, i.e.
\begin{align}
\zeta_r(x,0) &=  {\mathrm e}^{-36\left(\frac{4x}{L}-1\right)^{100}}\sum_{k>0} \left({\frac{|k|}{4\, g}}\right)^{1/4} \,\left ( a_k(0){\mathrm e}^{ikx}+a_k^*(0){\mathrm e}^{-ikx}\right)\,,\\
\zeta_l(x,0) &=  {\mathrm e}^{-36\left(\frac{4x}{L}-3\right)^{100}}\sum_{k<0} \left({\frac{|k|}{4\, g}}\right)^{1/4} \,\left ( a_k(0){\mathrm e}^{ikx}+a_k^*(0){\mathrm e}^{-ikx}\right)\,,\\
\phi_r(x,0) &=  {\mathrm e}^{-36\left(\frac{4x}{L}-1\right)^{100}}\sum_{k>0} -i \left({\frac{g}{4\, |k|}}\right)^{1/4}\,\left ( a_k(0){\mathrm e}^{ikx}-a_k^*(0){\mathrm e}^{-ikx}\right)\,,\\
\phi_l(x,0) &=  {\mathrm e}^{-36\left(\frac{4x}{L}-3\right)^{100}}\sum_{k<0} -i \left({\frac{g}{4\, |k|}}\right)^{1/4}\,\left ( a_k(0){\mathrm e}^{ikx}-a_k^*(0){\mathrm e}^{-ikx}\right)\,,
\end{align}
where the subscripts $r$ and $l$ are the right and left propagating packets respectively and $a_k(0)$ are the periodic initial condition Fourier coefficients. The full initial condition is thus simply $\zeta=\zeta_r+\zeta_l$ and $\phi=\phi_r+\phi_l$, see upper panel of figure \ref{fig:eta}. The solution is integrated until $T=100s$ which is roughly the time it takes the fastest travelling waves to traverse the $L=50m$ numerical tank.

We perform the same parameterised sweep of $\alpha$ as in Section \ref{sec:5mode} to intersect the resonance curve and seek increased transfer efficiency. We show the usual $\mathcal{E}$ efficiency for consistency with the previous sections. A slight adjustment is made since the smoothing envelope broadens the initial spectrum, we also allow for a similar widening of 10 adjacent modes in the transfer efficiency definition:

\beq
\label{eq:eff2}
 \mathcal{E}_{k}(t) = \frac{\sum\limits_{k'=k-10}^{k+10}\omega_{k'} a_{k'}a_{k'}^*}{H}\,.
 \eeq

We also consider an alternative estimate of the energy transfer which may be more experimentally practical. A high fidelity simple measure often made of such a system is a sampling of the free surface at a given location, e.g. at the centre of the domain $x=25 m$. If the precession resonance has resulted in significant energy transfer into mode $K_3,$ say, this would be manifest as the appearance of $\omega_3$ in the signal of $\zeta(x=25,t).$ We compute an appropriate measure of this by the Fourier integral

\beq
\label{eq:eff3}
 \mathcal{E}_{\omega}(x) = \frac{\int_{t_0}^T \zeta(x,t)\mathrm{e}^{\mathrm i \omega t}\mathrm{d}t}{\int_{t_0}^T |\zeta(\bar{x},t)|\mathrm{d}t}\,,
  \eeq
{i.e. the frequency content of a surface elevation measurement taken at $x$ over the time interval $t\in[t_0,T]$ and normalised by the integral of the surface elevation at  $x=\bar{x} \equiv 25m.$ For reasons of computational efficiency and accuracy, we take the time interval $[75,100]$ to capture the period of strongest growth when the target frequency is most evident. }

\begin{figure}
\begin{center}
\hspace{-5mm}
\scalebox{0.4}{\input{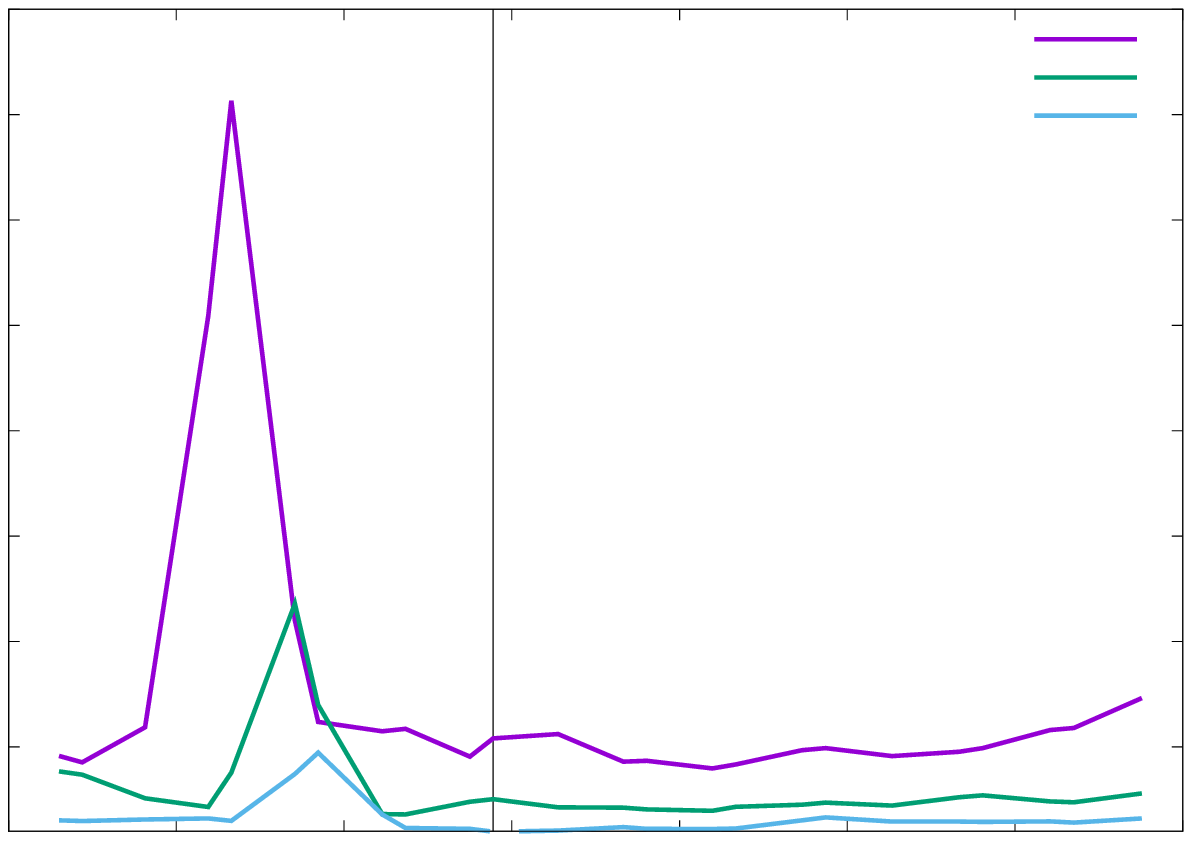}}
\hspace{5mm}
\scalebox{0.4}{\input{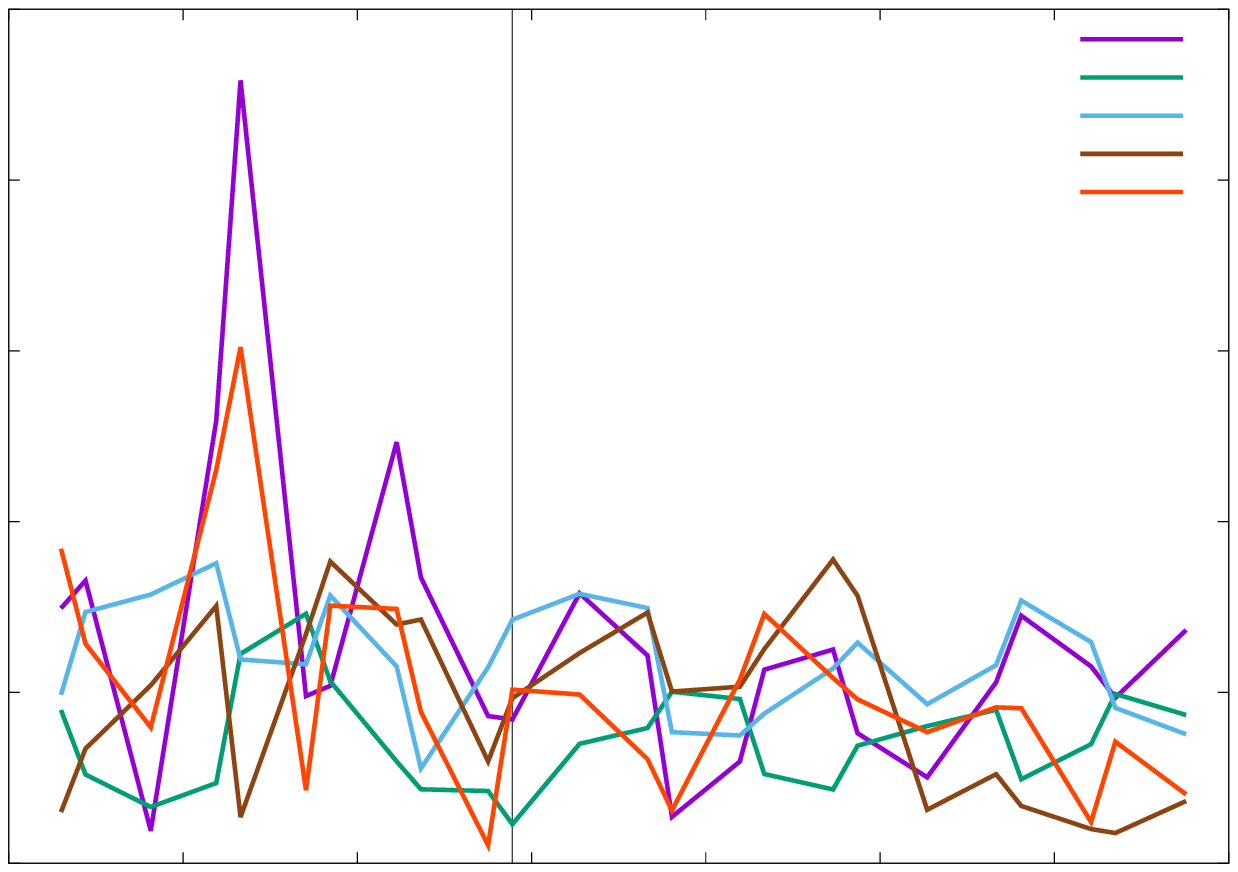}}\\
\vspace{-15mm}
\scalebox{0.4}{\input{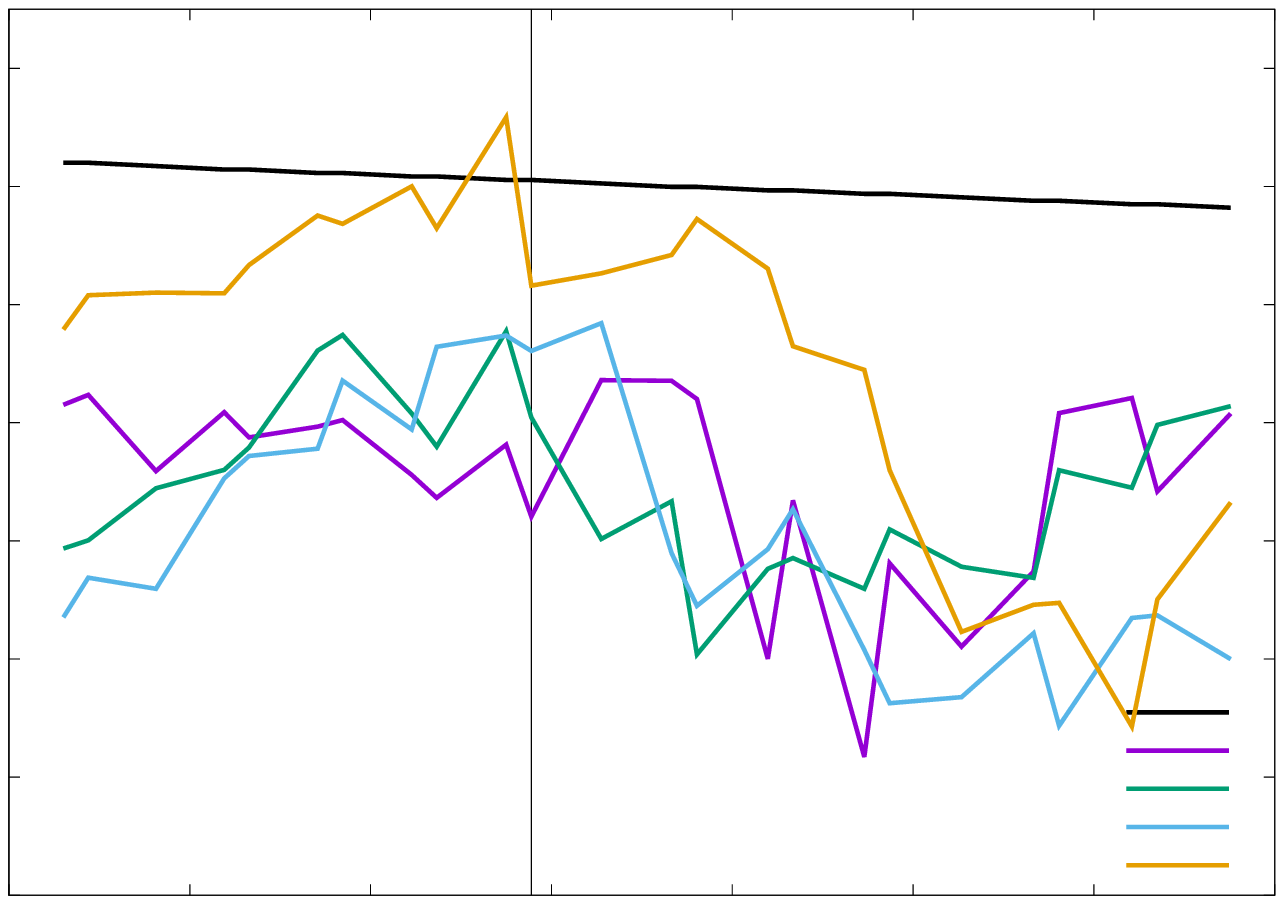}} \hspace{5mm}
\scalebox{0.4}{\input{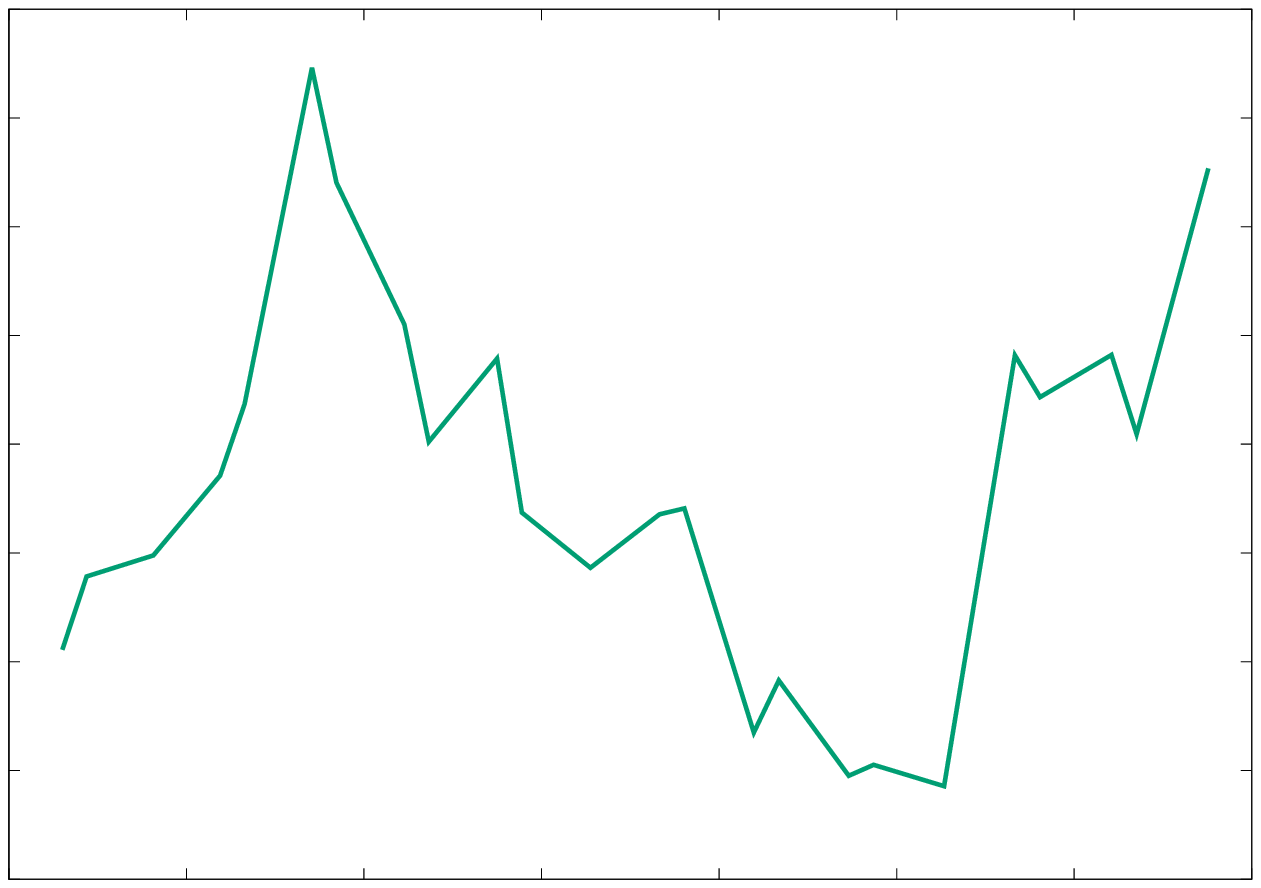}}
\vspace{-5mm}
\end{center}
\caption{Efficiencies $\mathcal{E}_{K_1}$ and $\mathcal{E}_{\omega_3}$ (upper), and precession $\Omega$ (lower left) plotted for three values of amplitude rescaling. Both indicate an enhanced transfer to the target modes at $\alpha\approx1.745.$ For $\mathcal{E}_{\omega_3}$ measurements offset from the centre of the tank, namely at $x=20m$ and $30m$ show that, as predicted by the theory, the transfer is into $K_3$ and not $-K_3$. Lower right presents the timescale $T_r=\frac{2\pi}{\langle | \Omega-2\omega_2| \rangle}$ showing a peak near the resonant peak of energy transfer. \label{packet1}}
\end{figure}

\begin{figure}
\begin{center}
\scalebox{0.4}{\input{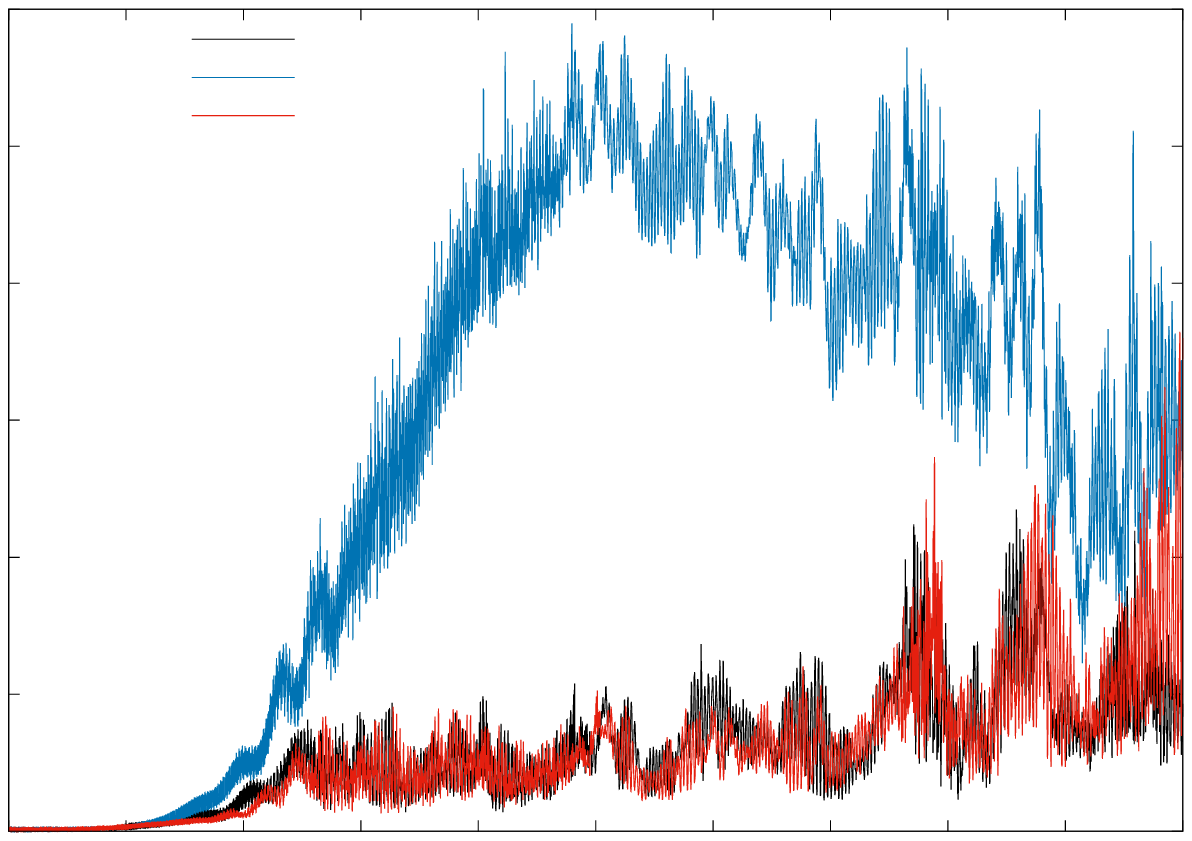}}
\scalebox{0.4}{\input{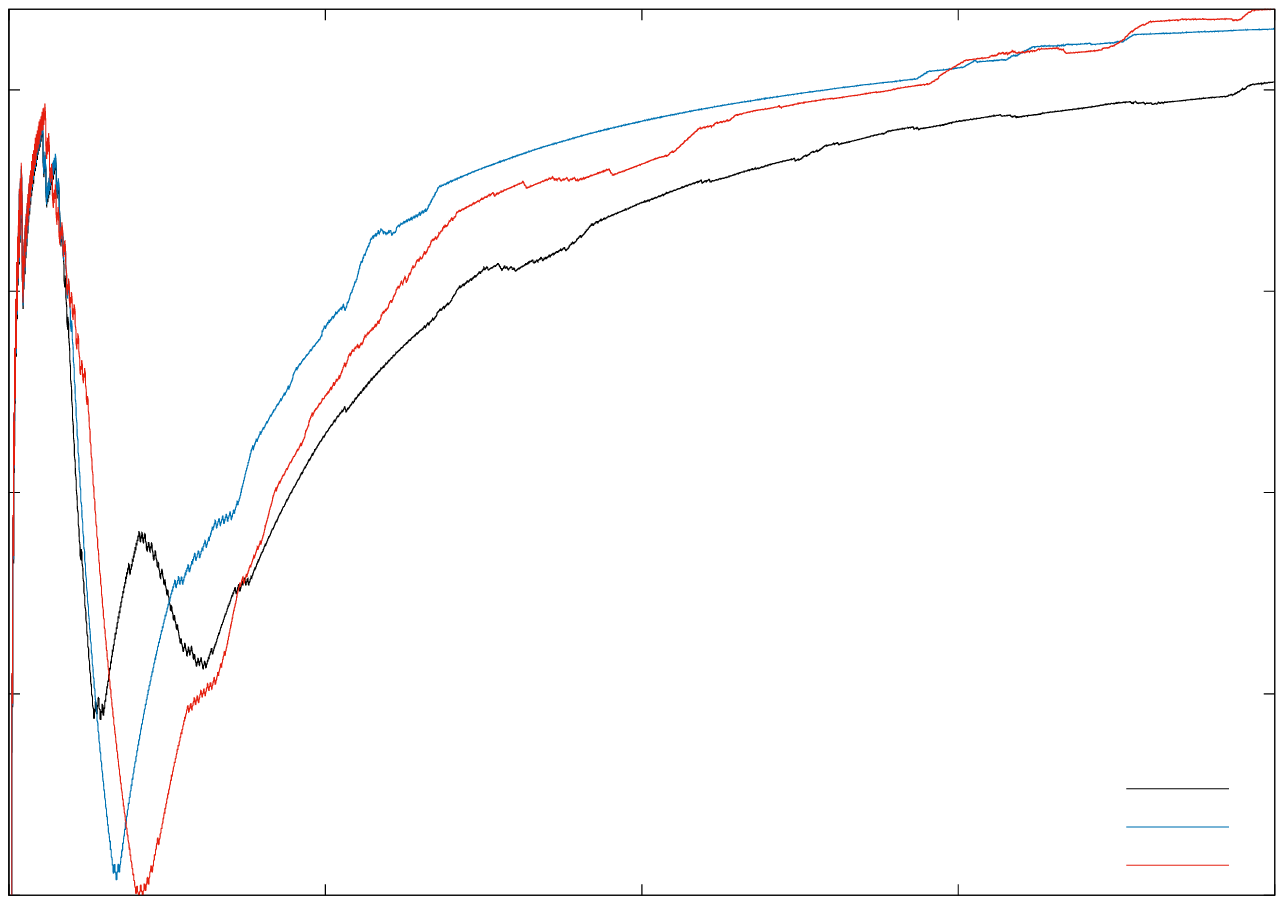}}
\vspace{-10mm}
\end{center}
\caption{ Efficiency (left) and precession $\Omega$ (right) plotted over time for three initial $\alpha$ values and $A=0.9.$ Notice the localised temporal window of energy transfer and slow convergence of precession. \label{packett}}
\end{figure}

Figure \ref{packet1} shows the efficiencies $\mathcal E_{-K_1}$ and $\mathcal E_{\omega_3}(25)$ for amplitude rescaling factors $A=1, \,\, 0.9$ and $0.8,$ and $\mathcal E_{\omega_3}(20)$, $\mathcal E_{\omega_3}(30)$ with $A=1.$ We see a similar increase of the peak efficiency in both efficiency measures with $A$ and a shift toward lower $\alpha$ as in the previous cases, although there is increased variability in the frequency measure $\mathcal E_{\omega_3}$. {In the offset probes we see small but appreciable resonant transfer at $\mathcal E_{\omega_3}(30)$ but no clear transfer at $\mathcal E_{\omega_3}(20),$ confirming that there is resonant transfer to $K_3$, and no transfer to $-K_3.$} The precession values do not show such a strong correspondence with the efficiency peaks and exhibit significant variations. 
Notice that the timescale of growth of energy in the target mode is generally longer than the 100-second limit we impose for these wave-packet experiments and is partially responsible for the weaker growth of energy. In particular the precession is not well converged in this short time window. Figure \ref{packett} shows the evolution of $\mathcal{E}_{K_3}$ and a running estimate of  $\Omega$ of a longer calculation with $T=200s.$ One can plainly see that the precession has not yet converged and is subject to significant variations.  Figure \ref{packet1} has included the precession value for this longer calculation which arguably gives a closer correspondence to the efficiency peak observed.
Similar to the broadening of the efficiency definition (\ref{eq:eff2}) one can also consider an average resonance condition across the set of triads generated by the ten nearest modes. In particular a resonance timescale defined as
$$ T_{r}=\frac{2\pi}{\langle |\Omega-2\omega_2|\rangle},$$
where angle brackets here denote an average over triads, will be expected to have a peak near the peak energy transfer. {Physically, $T_r$ is the maximal time scale at which we could possibly expect to see a sustained growth of the modes involved in the triads studied, in analogy to maximal timescale (\ref{eq:Tmax}) for a single triad.} Figure \ref{packet1} shows $T_r$ over the sweep of $\alpha$ and indeed shows a peak near the corresponding peaks of $\mathcal{E}.$ There is some small discrepancy which may be accounted for by the corrections to $\Gamma$ which have shown to become influential at larger nonlinearity (figure \ref{5mode1} lower right plot).

The time series of $\mathcal{E}_{K_3}(t)$ also highlights that the nonlinear interaction is confined to the central region where the two wavepackets are coincident; growth is strongest between $t=50$ and $100$ seconds and therefore the overall available energy to be exchanged is less than these {earlier} bulk measures considered. A quantitative comparison with the results of the planar cases of the previous sections is therefore not particularly helpful, suffice it to say that the resonance is observed.

\section{Discussion}

This work has shown for the first time the presence and importance of precession resonance in nonlinear surface gravity waves in an infinitely deep fluid. We have been able to show that, for reduced models, energy transfers can be enhanced by orders of magnitude when the resonance conditions are satisfied, and are strongly dependent on initial amplitude. Furthermore we have demonstrated that this effect should be observable in an experimental system involving encountering wave packets. {Moreover, a fundamental question is raised by the existence of the precession resonance mechanism: \emph{How to reconcile the weakly-nonlinear theory with this new mechanism?} In particular, we demonstrated that the timescale of precession resonance is inversely proportional to the wave amplitude, whereas the weakly-nonlinear theory predicts inverse proportionality to the third power of the wave amplitude. In future work we will tackle this important problem.}

Our evidence suggests that a domain {larger than $50$ metres} may be necessary to fully appreciate the precession resonance transfer mechanism using localised wave packets to allow longer times of interaction. However in a real physical laboratory system dissipation is present, in particular due to wave breaking. Here we have imposed no numerical dissipation method, principally to allow continuity from the low dimensional modelling to the wave packets but also for ease of implementation. As such the physical laboratory experiments would be able to attempt larger initial amplitudes which, given the evidence in Section \ref{sec:5mode}, would lead to faster energy transfer timescales and therefore larger overall efficiencies.   {We have performed preliminary experiments that seem to confirm the precession resonance mechanism, and will present a thorough account of experimental results in a subsequent paper.}

In addition we have a periodic domain and halt our simulation before the waves cross the domain boundary. In the physical tank waves could in principle be allowed to reflect off vertical walls if wavemakers could be lifted from the ends of the tank. This would allow the interactions to continue across larger spatial and temporal extents, again giving larger transfer efficiencies. One would hope that secondary interactions would not interfere with the energy exchanges under consideration here. 

Finally we have not exhaustively searched across all possible initial conditions, or even other resonances in {$K_1,$ $K_2$ space. A preliminary investigation to this end resulted in our presentation of $K_1/K_2=\alpha_c=16/9$} as a good candidate but others will exist and may show improved transfers in such conditions. A topic for future research is how best to isolate the most promising or indeed relevant candidate resonances.
In a similar vein we have shown how to tune the frequencies in the system via adjusting the wavelengths of the initial condition to trigger the resonance. It would also be possible with a finite depth dispersion relationship to, for example, vary the depth of the water column to tune the frequencies and thus trigger the resonance. 

{One of the motivations to study one-dimensional propagation of gravity water waves in the deep-water limit, was to demonstrate clearly the existence of precession resonance in a system where it is generally known that triad interactions do not play an important role. Having demonstrated that triads are essential even in {pure} gravity waves, we will devote future work to: (i) consider two-dimensional propagation of water waves; (ii) include capillarity, where triads are generally known to be important; (iii) study other kinds of wave systems (internal, inertial, etc.) which are based on triad interactions; and (iv) consider systems where quartets {(not triads)} are the lowest-order nonlinearity, {as in nonlinear Schr\"odinger equation,} and demonstrate that these systems are also able to exhibit precession resonance, {this time via quartet interactions.}}

Despite concentrating here on free-surface gravity water waves, and previously on Rossby/drift waves \citep{Bustamante:2014jf}, there is scope for this resonance to be important in a great many turbulent systems. While study is most straightforward when waves are dispersive and the system size small, it has been shown that even when this is not the case, precession behaviour is important at organising the dynamics \citep{Buzzicotti:2016bz}. {It is known that turbulent cascades are accompanied with bursts and intermittency, which require some amount of synchronisation of the phases of the modes involved in the transfers \citep{holmes1998turbulence,perri2012phase}. Also, many (if not all) of the mechanisms proposed for rogue wave generation require or imply some amount of phase coherence/synchronisation/constructive interference \citep{onorato2001freak,slunyaev2010freak,Akhmediev2010,grimshaw2013rogue,fedele2016real}. But our new mechanism deals with the \emph{precession frequencies} of the phases rather than the phases themselves, and we have established \citep{Bustamante:2014jf} that these precession frequencies do synchronise when strong collective energy-transfer events occur in the system, even when the phases do not synchronise. Therefore, in the near future we will focus our efforts to find applications of the precession resonance mechanism on cascades and rogue waves in general turbulent systems.}

\section*{Acknowledgements}
This research was financially supported by a research grant from Science Foundation Ireland (SFI) under grant no. 12/IP/1491. Computational resources were provided by the Irish Centre for High-End Computing (ICHEC) via class C project ndmat025c.

\bibliography{papers_DL,papers_MB}
\bibliographystyle{jfm}

\end{document}